# MN$_4$ Embedded Graphene Layers: Tunable Decay Rate of RKKY Interaction

Mahnaz Rezaei[1], Jahanfar Abouie[2,*], Fariba Nazari[1,3,†]

[1]Department of Chemistry, Institute for Advanced Studies in Basic Sciences (IASBS), Zanjan 45137-66731, Iran
[2]Department of Physics, Institute for Advanced Studies in Basic Sciences (IASBS), Zanjan 45137-66731, Iran
[3]Center of Climate Change and Global Warming, Institute for Advanced, Studies in Basic Sciences (IASBS), Zanjan 45137-66731, Iran

One of the most important tasks in the development of high-performance spintronic devices is the preparation of two dimensional (2D) magnetic layers with long-range exchange interactions. MN$_4$ embedded graphene (MN$_4$-G) layers, with M being transition metal elements, are experimentally accessible 2D layers, which exhibit interesting magnetic properties. In this paper, by employing the spin-polarized density functional theory (SP-DFT), we study MN$_4$-G layers with a special focus on the behavior of the indirect M-M exchange interactions, and demonstrate that the MN$_4$-Gs with M = Fe, Mn and Co, are 2D anisotropic magnetic layers with Ruderman-Kittel-Kasuya-Yosida (RKKY) interaction. By examining the electronic configurations of the M atoms for various M-M spacers, we demonstrate that the RKKY interaction in such layers are tunable and exhibiting an unusual prolonged decay ($r^{-n}$, with $0.5 \leq n \leq 2$). In addition, we investigate the influence of the CuN$_4$ moiety in the graphene host, and show that, in contrary to the other MN$_4$-Gs, the 2D CuN$_4$-G layer behaves as decoupled one-dimensional spin chains, regardless of the spacer lengths.

## I. INTRODUCTION

In the design of miniaturized spin-based devices, magnetic properties of the nanostructures that make up the device play a crucial role. A range of magnetic features, such as magnetotransport and overall magnetic moment formation, are predicated upon magnetic interactions and understanding the spin coupling within and between magnetic center (s) is of crucially importance to realize magnetic properties of materials [1-5].

The RKKY interaction, a conduction-electron-mediated interaction between magnetic atoms, plays a key role in the field of spintronic. This interaction is unique in its flexibility to tailor the physical properties of spin systems in terms of coupling strengths and noncollinearity [6]. The RKKY interaction is long-range, but decays by increasing the separation distance ($r$) of magnetic centers, as $r^{-n}$, where $n$ shows the interaction decay rate. Fast decay rate ($n > 2$) presents an obstacle to spintronic applications [7-25]. Any method for amplifying the coupling to extend its range is useful for both the experimental detection of the RKKY interaction and future spintronic applications. In this respect, extensive efforts have been devoted to fabricate desirable spin systems with RKKY interaction with slower decay rates [19-31].

2D materials in a precisely designed order can combine the best of different components in one ultimate material [32-35], and possess essential requirements for spintronic applications. Magnetic properties of different 2D layers arise from their embedded localized magnetic moments [32,36]. Among different 2D materials, graphene is a highly promising material for next-generation spintronic applications, owing to its extraordinary carrier mobility, long spin diffusion length, weak intrinsic spin–orbit coupling, and limited hyperfine interactions [37,38]. In addition, for practical applications in spintronic devices, the graphene-based material is required to be magnetic above room temperature. Various strategies have been attempted to realize magnetic ordering in graphene [15,39-52]. Recently, Hu et. al. reported a stable room temperature ferromagnetic ordering in graphene by embedding single magnetic Co atoms in the lattice via the strong chemical bonds in the CoN$_4$ moieties [53]. The stable room temperature planer graphene magnets developed in [53] do not have many disadvantages and obstacles of the previously reported modified graphene surfaces [54,55]. Actually, the experimentally accessible MN$_4$ embedded graphene (MN$_4$-G) layers, with M being the transition metal elements, are promising candidates to reach a stable magnetization on nanoscale [53,56,57].

In spite of accomplishment of long-range magnetic order in a purely 2D layer that has been actively pursued in recent years [1,48,58-60], there is no comprehensive information for the kind and the range of M-M exchange interactions in MN$_4$-G layers. In this paper, we determine the exchange interaction between magnetic atoms in the MN$_4$-G layers, by using the SP-DFT and Green's function methods [61-63] with the formulation of Wannier functions [64,65]. We study various MN$_4$-G layers with M=Fe, Co, Mn, Cu, and demonstrate that the studied MN$_4$-G layers can be categorized into two different groups. The first one includes layers where the RKKY interaction has unusual prolonged decay ($r^{-n}$, with $0.5 \leq n \leq 2$) in spite of the fast decay rate ($n > 2$) in most artificial 2D layers [7-25]. The slow decay rate of the RKKY interaction, which makes the MN$_4$-G layers suitable for applications, is added to the other unique features [66-71] of these experimentally accessible 2D layers. In the second group that are containing layers with Cu atoms, an interesting characteristic is the formation of decoupled one-dimensional spin chains, regardless of the kind of the M-M spacer. This feature makes the CuN$_4$-G layer desirable to see different interesting phases like the Tomonaga–Luttinger liquid phase [72], experimentally.

* jahan@iasbs.ac.ir
† nazari@iasbs.ac.ir

The structure of this paper is as follows: In Sec. II, we introduce computational methods and model systems for computing exchange interaction parameter ($J_{ij}$) and orbital contribution to the total exchange parameter. In Sec. III, we tackle the exchange interaction between magnetic moments on the MN$_4$-G 2D layers and shed light on the exchange mechanism. To this end in Sec. IIIA we have analyzed the electronics and magnetic properties of the studied layers and in Sec. IIIB by analyzing the obtained data we discuss the exchange interaction mechanisms. Finally, in Sec. IV we summarize the main findings of our study.

## II. COMPUTATIONAL METHODS

We have employed quantum espresso code [73] to carry out structural relaxation of MN$_4$ (M=Fe, Co, Mn, Cu) embedded graphene (MN$_4$-G) layers by the spin-polarized density functional theory (SP-DFT) based methods [74].

The supercell models of layers in the present study are based on graphene monolayer with different supercell sizes. We constructed three distinct layers of various sizes (R1, R2, and R3) that have different MN$_4$ concentration and different spacer lengths between M atoms (See Sec. III).

To eliminate the interlayer interaction, a vacuum distance of ~10 Å between adjacent layers is adopted in all calculations. Here, we adopt the Perdew–Burke–Ernzerhof revised for solids (PBEsol) [75] XC functional within the generalized-gradient approximation (GGA) [76].

The ion−electron interaction is treated by the projector-augmented-wave (PAW) method [77]. Sampling of the Brillouin zone, a Γ-centered Monkhorst–Pack $k$-point mesh of 2×2×1 and 5×5×1 is used for different layers [78], and an energy cutoff of 1088.46 eV is adopted for the plane wave basis. The structure optimization stopped when the residual force on each atom was less than 0.026 eV/Å and energy was smaller than 0.014 eV. We have calculated the Hubbard parameter using hp.x. It is based on density functional perturbation theory (DFPT) [79]. The calculated Hubbard parameter and structural parameters for all studied layers reported in Tables I and II, respectively.

To investigate the stability of the layers we rely on the cohesive energy defined as:

$$E = \frac{[E_{MN_4-G} - (N_C E_C + N_N E_N + N_M E_M)]}{q}, \quad (1)$$

where $E_{MN_4-G}$ is the total energy of the studied layers, and $E_C, E_N$ and $E_M$ are the total energies of free C, N, and M atoms, respectively, and $N_C, N_N$, and $N_M$ are the number of C, N, and M atoms in the supercell, respectively, and $q$ stands for the total number of atoms in the supercell. The cohesive energies of the studied layers are reported in Table I. These values clearly indicate that the cohesive energy of the MN$_4$-G layers is in the range of the cohesive energy of graphene. A deep fundamental knowledge of the structure–electronic property relationships, which are difficult to access from experiments, pave the way to determine the size and chemical environment dependency of the exchange interactions and related mechanisms. By using these data, we tackle the exchange interaction between magnetic moments on the MN$_4$-G 2D layers.

One of the most common and successful microscopic models for describing magnetic properties of different systems is the Heisenberg model with localized spins:

$$H = \sum_{i \neq j} J_{ij} S_i \cdot S_j, \quad (2)$$

where $J_{ij}$ is the exchange coupling of the spins sitting at $i$ ($S_i$) and $j$ ($S_j$). There exist a few methods to evaluate $J_{ij}$ within SP-DFT and map the obtained results onto the Heisenberg model. The most direct, and common way to calculate $J_{ij}$ is to calculate the total energies of the $P + 1$ magnetic configurations, where $P$ is the number of different exchange interaction [80-82]. Despite the robustness of this approach, it has several serious drawbacks. (1) for complicated systems a number of different magnetic configurations have to be calculated; (2) all configurations should have the same magnetic moments; and (3) the result is hard to analyze, since it is purely a number, i.e., understanding the orbitals contribution in exchange interaction [83]. To overcome these shortcomings the Green's function method [61-63] is suggested. For the changes in the total energy with respect to small spin rotations, using SP-DFT and Heisenberg model, it produces analytical expressions. By adapting this approach, one not only obtain all the exchange interactions from the calculation of a single magnetic configuration, but also find contributions to the total exchange interaction coming from different orbitals.

Green's function approach was formulated for localized orbitals methods, [84,85]. However, most of modern band structure calculations are based on the methods, which use a plane-wave-type basis. They are the full-potential (linearized) augmented plane-wave (L)APW [86] and pseudopotential [87] methods. Consequently, within plane-wave approaches, a straightforward realization of the Green's function method becomes impossible. In addition, not all its advantages can be taken in the modern ab initio DFT based codes without direct definition of a localized basis set.

Dm. M. Korotin et. al. [83] adapted the Green's function approach for the plane-wave-based methods using the Wannier functions formalism [64,65]. Application of the local force theorem (see, e.g., Ref. [88]) is the power of the Green's function method. The resulting change in the total energy due to the spins rotation over a small angle can be calculated via the local force theorem [62]. These are possible if the Hamiltonian of the system is defined in a localized orbitals basis set. The result of the rotation is compared with a similar procedure performed for the spin-Hamiltonian Eq. (3), which allows deriving an analytical expression for the exchange integrals [83]. To remove the major difficulty in the application of this approach to the modern plane-wave-based calculation Dm. M. Korotin et. al. [83] propose to use the Wannier functions projection [64,65] procedure and show its realization for the pseudopotential method.

The transformation of plane-wave eigenstates to Wannier functions has carried out using the wannier_ham.x utility that is available within the quantum espresso code [73]. To obtain the exchange interactions in different MN$_4$-G layers, we considered generalized Heisenberg Hamiltonian:

$$H = \sum_{i \neq j} J_{ij} e_i e_j, \qquad (3)$$

where $e_i$ and $e_j$ refers to the spin unit vectors of atomic sites $i$ and $j$. A negative (positive) sign of $J_{ij}$ indicates a ferromagnetic (antiferromagnetic) interaction. It should be noted that each $J_{ij}$ is now expressed as the following matrix whose elements represent the interactions between two given orbitals, and the conventional value of $J_{ij}$ is given by the sum of all these matrix elements [89]:

$$J_{ij} = \begin{pmatrix} d_{z^2} d_{z^2} & d_{z^2} d_{xz} & d_{z^2} d_{yz} & d_{z^2} d_{x^2-y^2} & d_{z^2} d_{xy} \\ d_{xz} d_{z^2} & d_{xz} d_{xz} & d_{xz} d_{yz} & d_{xz} d_{x^2-y^2} & d_{xz} d_{xy} \\ d_{yz} d_{z^2} & d_{yz} d_{xz} & d_{yz} d_{yz} & d_{yz} d_{x^2-y^2} & d_{yz} d_{xy} \\ d_{x^2-y^2} d_{z^2} & d_{x^2-y^2} d_{xz} & d_{x^2-y^2} d_{yz} & d_{x^2-y^2} d_{x^2-y^2} & d_{x^2-y^2} d_{xy} \\ d_{xy} d_{z^2} & d_{xy} d_{xz} & d_{xy} d_{yz} & d_{xy} d_{x^2-y^2} & d_{xy} d_{xy} \end{pmatrix}. \qquad (4)$$

The MN$_4$-G layers studied in this work have high geometric anisotropy, which affects the exchange interaction between the M atoms. We have calculated the exchange interaction matrix (EIM) in different directions. These are $x$, $y$, and two main diameters (**d1**, **d2**) directions as shown in Fig. 1(a).

The convergence of $J_{ij}$ with respect to the number of $k$-points is assessed with a progressively finer mesh. We have used 10×10×1 and 16×16×1 $k$-points mesh for the studied layers.

These parameters produce $J_{ij}$ values in the range of the reported values for layers that are containing similar elements [90].

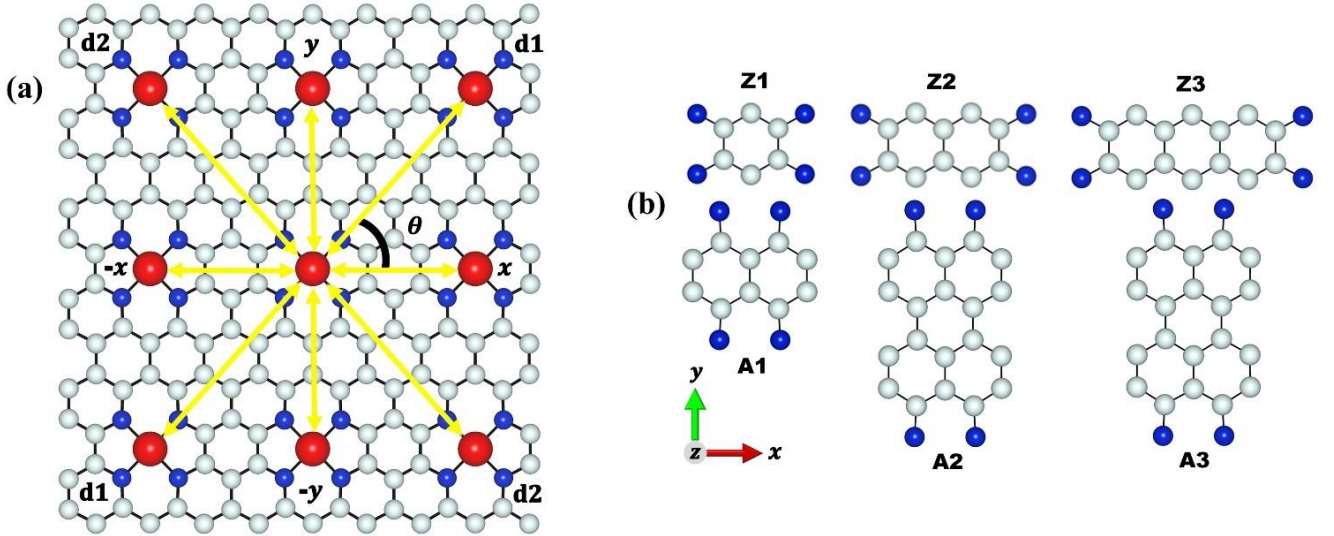

**FIG. 1.** (a) The exchange interaction of the M atom located at the origin with the neighboring atoms are shown with the yellow double head arrows along the $\pm x, \pm y$, and the two main diameters (**d1**, **d2**) directions. $\theta$ is its angle with $x$-axis. (b) Spacers with three sizes in $x$ and $y$ directions. Z3 (A3), Z2 (A2) and Z1 (A1) are spacers with zigzag (armchair) edges along the $x$ ($y$) direction in MN$_4$-G (R3), MN$_4$-G (R2) and MN$_4$-G (R1) layers, respectively.

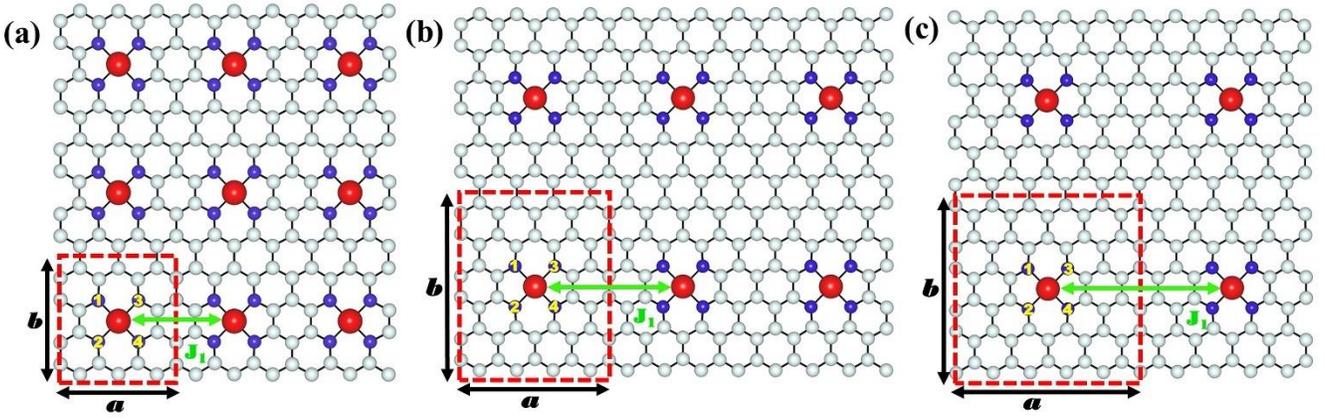

**FIG. 2.** Geometry of different $MN_4$-G (M=Fe, Co, Mn, Cu) layers, (a) R1, (b) R2, (c) R3. The gray, blue and red balls stand for C, N and M atoms, respectively. $J_1$ is the exchange interaction between nearest-neighbor M atoms.

**Table I.** Cohesive energy (CE), Hubbard parameter (HP) and nearest-neighbor exchange interaction ($J_1$) for $MN_4$-G (R1, R2, R3) layers.

| Layer | CE/ atom (eV) | HP (eV) | $J_1$ (meV) | Layer | CE/ atom (eV) | HP (eV) | $J_1$ (meV) |
|---|---|---|---|---|---|---|---|
| Graphene | -5.83 | --------- | --------- | --------- | --------- | --------- | --------- |
| $MnN_4$-G (R1) | -7.48 | 5.9008 | -1.748 | $CoN_4$-G (R1) | -7.61 | 6.7809 | -0.825 |
| $MnN_4$-G (R2) | -7.90 | 6.0171 | 0.054 | $CoN_4$-G (R2) | -7.96 | 7.0114 | 0.054 |
| $MnN_4$-G (R3) | -7.97 | 5.9350 | 7.367 | $CoN_4$-G (R3) | -8.02 | 7.6818 | 0.001 |
| $FeN_4$-G (R1) | -7.48 | 6.4328 | -9.433 | $CuN_4$-G (R1) | -7.50 | 6.4180 | 1.747 |
| $FeN_4$-G (R2) | -7.93 | 5.9513 | 0.076 | $CuN_4$-G (R2) | -7.91 | 6.4002 | 0.190 |
| $FeN_4$-G (R3) | -8.00 | 5.8474 | 0.049 | $CuN_4$-G (R3) | -7.98 | 6.4303 | 0.018 |

**Table II.** $a$ (distance between the M atoms in the $x$ direction) and $b$ (distance between the M atoms in the $y$ direction) are lattice parameters. Structural parameters are $d$ [(M-$N_{1(2,3,4)}$), bond-lengths], F [($N_{1(3)}$-M-$N_{2(4)}$), bond-angles, (degree)] and G [($N_{1(2)}$-M-$N_{3(4)}$), bond-angles, (degree)]. $m$ is effective magnetic moment of M atom per cell ($\mu_B$/cell). The $MN_4$-G (R1, R2, R3) layers are shown in Fig. 2(a-c) and the four N atoms are numbered.

| Layer | $a$ (Å) | $b$ (Å) | $d$ (Å) | F | G | $m$ | Layer | $a$ (Å) | $b$ (Å) | $d$ (Å) | F | G | $m$ |
|---|---|---|---|---|---|---|---|---|---|---|---|---|---|
| $MnN_4$-G (R1) | 7.49 | 8.29 | 1.86 | 91.27 | 88.73 | 3.00 | $CoN_4$-G (R1) | 7.46 | 8.25 | 1.84 | 90.69 | 89.31 | 1.00 |
| $MnN_4$-G (R2) | 9.92 | 12.61 | 1.87 | 90.86 | 89.14 | 3.00 | $CoN_4$-G (R2) | 9.89 | 12.57 | 1.84 | 90.73 | 89.28 | 1.00 |
| $MnN_4$-G (R3) | 12.38 | 12.64 | 1.87 | 90.62 | 89.39 | 3.50 | $CoN_4$-G (R3) | 12.36 | 12.62 | 1.85 | 90.63 | 89.36 | 1.00 |
| $FeN_4$-G (R1) | 7.48 | 8.26 | 1.84 | 90.77 | 89.23 | 3.76 | $CuN_4$-G (R1) | 7.49 | 8.30 | 1.89 | 90.97 | 89.03 | 1.16 |
| $FeN_4$-G (R2) | 9.91 | 12.58 | 1.85 | 90.60 | 89.40 | 2.00 | $CuN_4$-G (R2) | 9.92 | 12.61 | 1.90 | 90.96 | 89.04 | 1.00 |
| $FeN_4$-G (R3) | 12.37 | 12.63 | 1.86 | 90.61 | 89.39 | 2.00 | $CuN_4$-G (R3) | 12.37 | 12.66 | 1.90 | 90.97 | 89.03 | 1.00 |

## III. RESULTS AND DISCUSSIONS

### A. ELECTRONIC AND MAGNETIC PROPERTIES

By considering supercells of various sizes, particularly supercells of 23, 47, and 59 atoms including one transition metal site (see Fig. 2(a-c), the red dashed boxes), we have studied the effects of different environments on the behavior of M-M exchange interaction. To this end, we first report our SP-DFT results on the ground state electronic and geometric properties of the $MN_4$-G layers, by clarifying the role of transition metal elements, the chemical-environment, the surface geometric anisotropy, and the transition metal concentration. Similar to the optimized graphene layers in which the transition metal doped vacancies create a Jahn-Teller distortion around the vacancy [91-93], all supercells are optimized to distorted structures in the presence of $MN_4$ moiety (Table II). Our results show that the optimized structures keep the planar structures without any wrinkle. These are in good agreement with the available optimized structure for some layers reported by different groups [67,90,94-96]. The three types of supercells that we have considered in our study possess intrinsic geometric anisotropy, and the M atoms in the $MN_4$-G layers are separated by different kind of spacers, i.e., spacers with the two different zigzag (Z) and armchair (A) edges, and with various lengths in $x$ and $y$ directions (R1, R2, and R3) (see Fig. 1(b)). The cohesive energies (Table I) clearly indicate the stability of the studied $MN_4$-G (R1, R2, R3) layers.

The different spacer lengths lead to different surface coverage of magnetic atoms, and consequently to different electronic and magnetic properties. To interpret the mechanisms of the exchange interactions, among various electronic properties we will focus on localization, delocalization, orbital contribution in electronic properties and bandwidth in the electronic bands. We recall that the bandwidth is a good measure of energy that is gained by electron hopping [97]. The bandwidth of the $MN_4$-G (R1) is larger than $MN_4$-G (R2, R3) layers (the yellow shadow area in Fig. 3). Therefore, in $MN_4$-G (R1) layers electrons can hop more likely. Another interesting feature seen in the band structure of $MN_4$-G layers (see Fig. 3) is the emergence of nearly dispersion-less bands, known as flat bands. As we know, the density of states really measures the number of states in a small energy range. Therefore, flatter bands will have more states over a small energy range in comparison with a highly dispersive band. In other words, the bandwidth (see Fig. 3) is related to the overlap between orbitals on nearest-neighbor atoms. Some of the $d$-orbitals of the M atom are almost localized and have less overlap with the neighboring N atoms. So, the flat bands correspond to the localized $d$-orbitals of the M atom similar to those of reported by Orellana and Wu [66,70]. The projected density of states confirms the almost localized nature and less overlap of the $d$-orbitals with neighboring N atoms (Appendix AI), too. The calculated spin-polarized band structure show that, in both spin channels all the $MN_4$-G (R2, R3) layers have a metallic character, while the other $MN_4$-G (R1) layers, except for $FeN_4$-G (R1), are semiconductor. The indirect band gap in the majority-spin channel (minority-spin channel) of the $MnN_4$-G (R2), $CoN_4$-G (R2), and $CuN_4$-G (R2) is 0.48 eV (0.11 eV), 0.40 eV (0.11 eV) and 0.25 eV (0.22 eV), respectively. The band gap of the minority-spin channel is much smaller than that of majority-spin channel (see Fig. 3(a, g, j)). These properties help us to illustrate the mechanism of exchange interactions, which we will discuss in the following section.

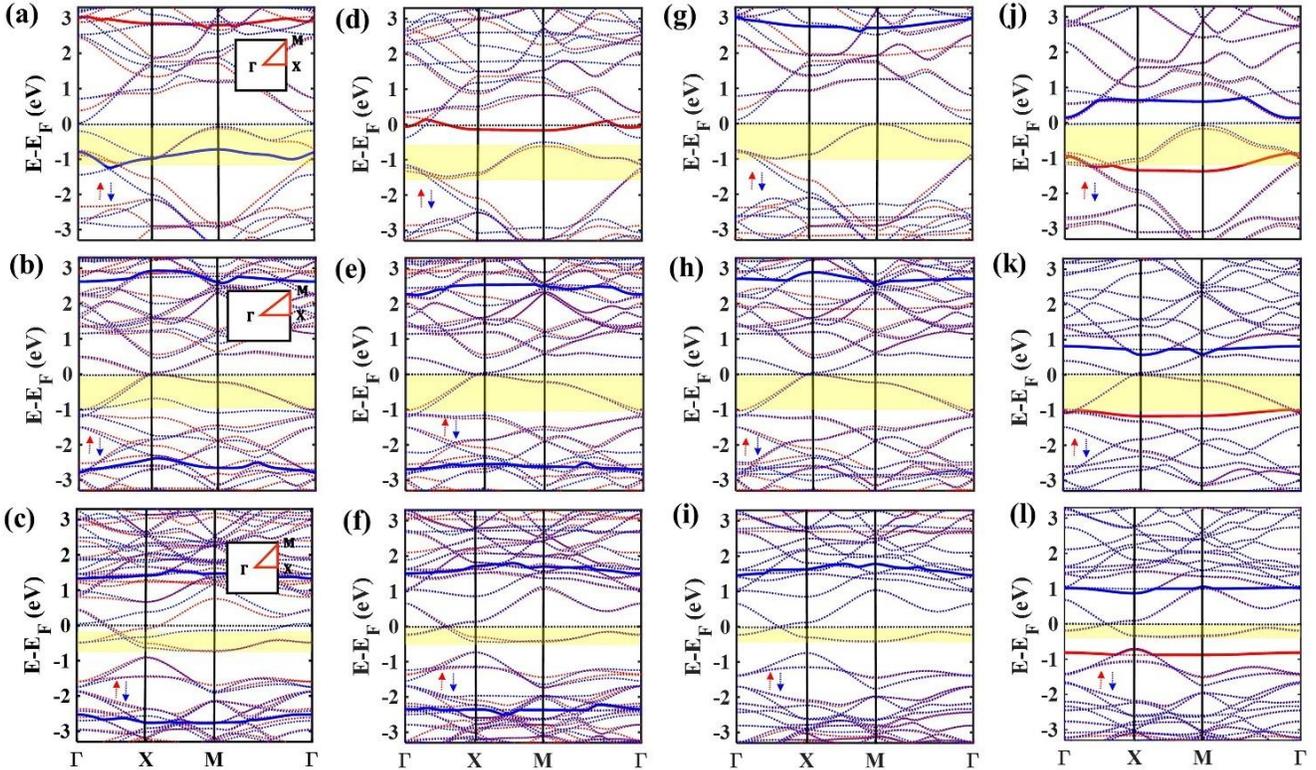

**FIG. 3.** Spin-polarized band structure of (a) $MnN_4$-G (R1), (b) $MnN_4$-G (R2), (c) $MnN_4$-G (R3), (d) $FeN_4$-G (R1), (e) $FeN_4$-G (R2), (f) $FeN_4$-G (R3), (g) $CoN_4$-G (R1), (h) $CoN_4$-G (R2), (i) $CoN_4$-G (R3), (j) $CuN_4$-G (R1), (k) $CuN_4$-G (R2), and (l) $CuN_4$-G (R3) layers. The yellow shadow area demonstrates bandwidth and the thick solid line shows the nearly flat bands. The Fermi level is set to zero. The inserted irreducible Brillouin zone in (a), (b) and (c) are for R1, R2 and R3 supercells, respectively.

There are different parameters, which are important in specifying the M-M exchange interactions [98]. The M-N bond-length is recognized as the main parameter to describe the chemical interactions and reflects the correlation of atomic-level configurations. We have compared the four M-N bond-lengths for different MN$_4$-G layers, in the Table II. As seen, all the bond-lengths are in the range of 1.84-1.90 Å. The minimum (maximum) of bond-length belongs to the Co–N and Fe-N (Cu–N). Furthermore, the ligand and the exchange fields are also important for specifying the effective magnetic moments ($m$) of the M atoms. In Fig. AII, we have shown their effects on the energy splitting of the $d$-orbitals. Although the different spacers have the same effects on the splitting of $d$-orbitals, the amount of splitting is different (see Fig. AII). In addition, except for the CuN$_4$-G, all the other studied layers show almost significant spin splitting (see Fig. 4).

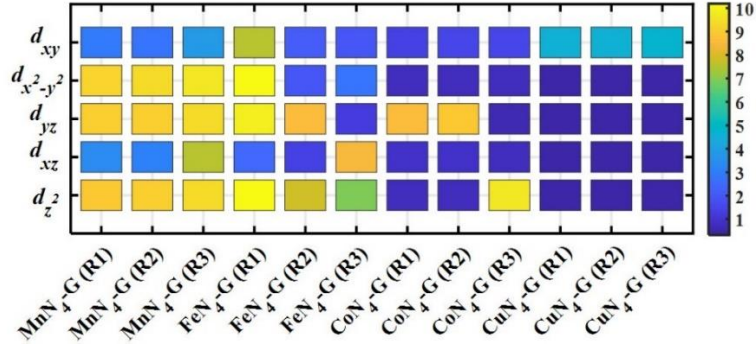

**FIG. 4.** Spin splitting (in eV) of $d$-orbitals of M atom in MN$_4$-G (R1, R2, R3) layers.

The N atoms around the M in the MN$_4$ moiety play a crucial role in determining the effective magnetic moment of the M atoms [99,100]. They cause the effective magnetic moment in most MN$_4$-G layers to be larger than those in the elemental solids (for bcc Fe, $m = 2.13$ $\mu_B$ [101], for alpha-Mn, depending on the local symmetry around the Mn, $m$ varies from 0 to 2.90 $\mu_B$ [102], and for fcc Cu, the system is diamagnetic and $m = 0$ [103]). The computed effective magnetic moments for MN$_4$-G (R1, R2, R3) in our study (Table II) are consistent with other available theoretical data [66,99,100]. The non-integer effective magnetic moments can be attributed to the strong interaction between M and its coordinating N atoms. Our results show that layers with the same effective magnetic moment have almost similar spin splitting trend (Fig. 4). Analysis of spin-polarized states above and below the Fermi level provides an estimate of the increase or decrease of the orbital occupation number. Consequently, nonzero magnetic moment is due to the difference in the orbital occupation numbers of majority- and minority-spin states. All majority-spin states are almost occupied, while the situation is different for minority-spin states. The layers with large (small) effective magnetic moments have minority-spin states with lower (higher) occupation numbers. To illustrate the detailed features of the effective magnetic moments of the MN$_4$-G (R1, R2, R3) layers, in Fig. AIII we have shown the spin density distribution. As seen, the spin density is highly localized on the transition metal atoms, and except for CuN$_4$-G (R1, R2, R3) layers, the spin density on the N atoms is small. Fig. 3 shows the spin-polarized band structures of the MN$_4$-G (R1, R2, R3) layers, including majority- and minority-spin sub-bands. Accordingly, these layers present different electronic properties that is consistent with the results reported in [70] for different supercell sizes of graphene.

### B. EXCHANGE MECHANISMS

As the studied layers are geometrically anisotropic, we calculate the exchange interaction matrix (EIM) elements (Eq. 4) of two M atoms in the $x$, $y$, **d1** and **d2** directions, (see Fig. 1(a)).

Our results show that in the $x$ and $y$ (**d1** and **d2**) directions the matrices are diagonal (non-diagonal), and the off-diagonal elements are proportional to the overlap of different orbitals, and the symmetries of the overlapping orbitals are very important. Moreover, all the EIM elements along the **d1** direction are equal to the corresponding ones along the **d2** direction with opposite signs.

To elucidate this, we have depicted in Fig. 5 the $d_{xz}$- and $d_{yz}$-orbitals of the M atoms, as an example. The overlap of the unlike lobes in the directions ($\mp x, \pm y$), (see Fig. 5, the red dashed arrows), and the like lobes in the direction ($\pm x, \pm y$), (see Fig. 5, the blue dashed arrows), results in the same amount with opposite signs for the corresponding off-diagonal elements of the EIM.

In the following, we compare the exchange coupling of the nearest-neighbor magnetic atoms ($J_1$), for different MN$_4$-G (R1, R2, R3) layers (see also Table I). The matrix elements of $J_1$ show the overlap of different $d$-orbitals of two M atoms along $x$ direction. In Fig. 6, we have reported the matrix elements of $J_1$ for the MnN$_4$-G (R1, R2, R3) layers (for the other studied layers, see Fig. BI.)

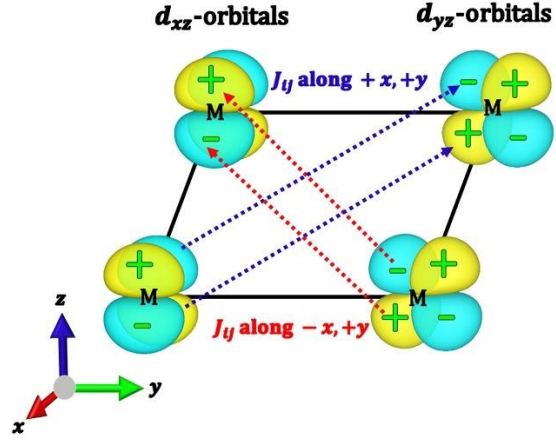

**FIG. 5.** Spatial orientations of the $d$-orbitals ($d_{xz}$ and $d_{yz}$) of the M atoms in MN$_4$-G (R1, R2, R3) layers. Positive (negative) orbital lobes are shown with the yellow (cyan) color.

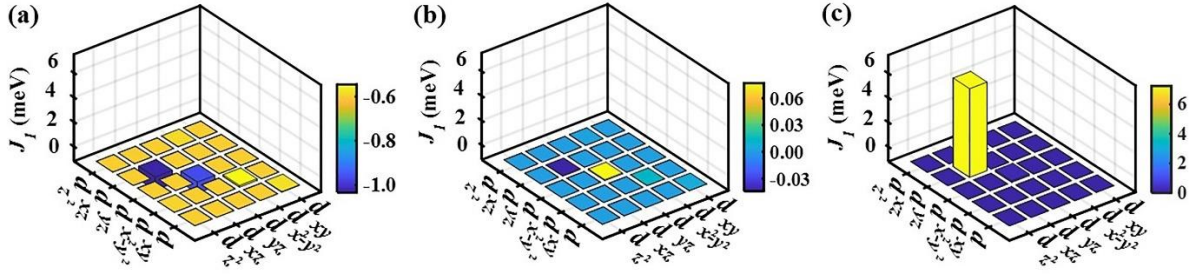

**FIG. 6.** The matrix elements of $J_1$ for (a) MnN$_4$-G (R1), (b) MnN$_4$-G (R2), and (c) MnN$_4$-G (R3) layers.

As it is clearly observed, in MnN$_4$-G (R2), FeN$_4$-G (R2, R3), and CoN$_4$-G (R1, R2, R3) layers, the $d_{yz}$-orbitals, in CuN$_4$-G (R1, R2, R3) and FeN$_4$-G (R1) layers, the $d_{xy}$-orbitals, and in MnN$_4$-G (R1, R3) the $d_{xz}$-orbitals have dominant role in specifying the exchange interaction $J_1$. Additionally, we should mention that only one of the diagonal matrix elements have significant contribution in each layer, and the off-diagonal elements are zero. We have observed that MnN$_4$-G (R3) and FeN$_4$-G (R1) are the only layers with substantially large $J_1$. It is ferromagnetic ($J_1 < 0$) in the MnN$_4$-G (R1), FeN$_4$-G (R1), and CoN$_4$-G (R1) layers, while antiferromagnetic ($J_1 > 0$) in the rest. As we have reported in the Table II, the distance between nearest-neighbor magnetic atoms is 7.48 Å for the FeN$_4$-G (R1) layer, and $J_1$ is -9.433 meV (see Table I). These are consistent with the corresponding ones reported in [90].

Now, let us investigate the behavior of the M-M exchange interaction ($J_r$, $r = x, y, \mathbf{d1}, \mathbf{d2}$) with respect to the M-M separation distance ($r$). Based on the dependence of $J_r$ to $r$, we have categorized the MN$_4$-G layers into two groups (Fig. 7). In the first group, which includes MnN$_4$-G (R1, R3), FeN$_4$-G (R1, R3) and CoN$_4$-G (R1, R3) layers, the exchange interaction $J_r$ decreases by increasing of $r$ and oscillates as it decreases (see Fig. 7(a, c, d, f, g, i)). In these layers, effective magnetic moments interact with each other via the long-range RKKY exchange interaction. This interaction occurs not via the direct $d$-$d$ exchange, which is usually small because the electrons' wave-functions are strongly localized in the $d$-orbitals, but rather through the delocalized conduction electrons of the $p_z$-orbitals of the N and C atoms. Since the spin polarization on N atoms is higher than C, the delocalized orbitals of the N atoms have more contribution to the RKKY interaction.

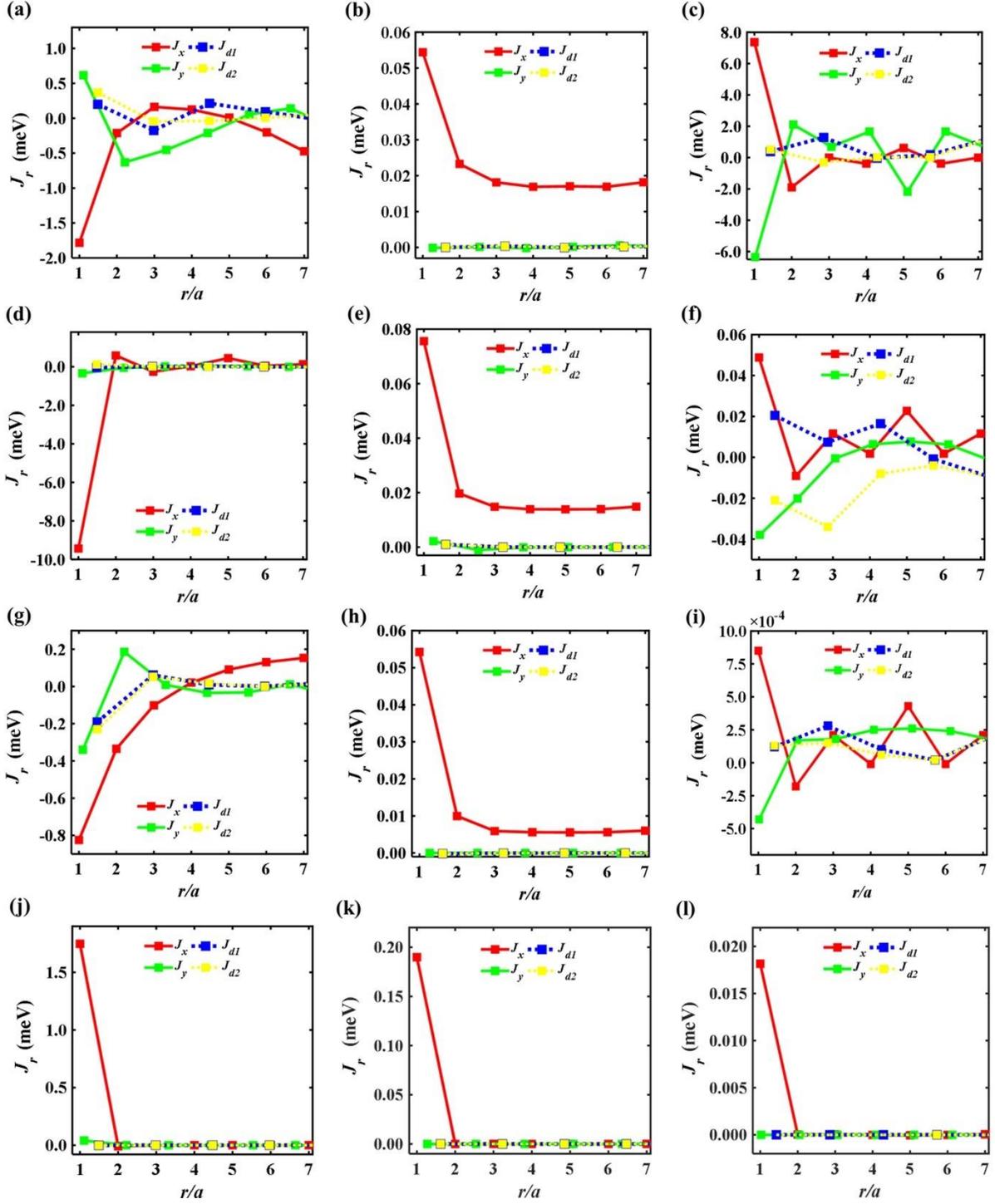

**FIG. 7.** Exchange interaction between two M atoms as a function of their distance ($r/a$) along x, y, **d1** and **d2** directions. The first, second and third columns are for two dimensional lattices with respectively R1, R2 and R3 supercells: (a) MnN$_4$-G (R1), (b) MnN$_4$-G (R2), (c) MnN$_4$-G (R3), (d) FeN$_4$-G (R1), (e) FeN$_4$-G (R2), (f) FeN$_4$-G (R3), (g) CoN$_4$-G (R1), (h) CoN$_4$-G (R2), (i) CoN$_4$-G (R3), (j) CuN$_4$-G (R1), (k) CuN$_4$-G (R2), and (l) CuN$_4$-G (R3) layers.

The strength of the RKKY interaction depends on the density of states of the conduction electrons at Fermi level, and on the overlap of the $d$-orbitals of the M atoms with conduction electrons [104]. As shown in Fig. AI(a, c, d, f, g, i), the $d$-orbitals with a large density of states close to the Fermi level have more contributions to the $J_1$ matrix. This can be understood from the charge density distribution near the Fermi level (see Fig. AIV(a, c, d, f, g, i)).

The second group includes the MnN$_4$-G (R2), FeN$_4$-G (R2), CoN$_4$-G (R2), and CuN$_4$-G (R1, R2, R3) layers. In these layers the exchange interaction is short-range and $J_r$ decays exponentially by increasing $r$. This occurs only in the $x$ direction, and the exchange interactions are negligible along the $y$, **d1**, and **d2** directions (Fig. 7(b, e, h, j, k, l)). The contributions of the $d$-orbitals of the M atoms near the Fermi level (see Figs. AI(b, e, h, j, k, l) and AIV(b, e, h, j, k, l)) are negligible, therefore, the exchange interaction mechanism between M atoms is of superexchange type.

In the second group, CuN$_4$-G is an exceptional case because in all (R1, R2, and R3) layers, $J_{y,d1,d2} \cong 0$, while $J_x$ is non-zero for nearest-neighbor Cu atoms. This means that the CuN$_4$-G layers can be effectively served as 1D antiferromagnetic spin-1/2 chains. As we know different interesting phases, like the Tomonaga–Luttinger liquid, are seen in the ground state phase diagram of the antiferromagnetic spin-1/2 chains. In this respect, all the studied CuN$_4$-G layers can be used as candidates for detecting such phases experimentally.

It is well known that the $d$-$p$ hopping integral and the energy difference between involved $d$- and $p$-orbitals are two major factors that determine the strength of superexchange couplings. To have an estimate for the measure of the $d$-$p$ hopping integral we have compared the integrated area of the unoccupied $d$-states of the MnN$_4$-G (R2), FeN$_4$-G (R2), CoN$_4$-G (R2), and CuN$_4$-G (R1, R2, R3) layers in the different energy intervals (see Fig. AV). The first three layers have similar integrated areas and the unoccupied $d$-states are more delocalized, and participated in hybridizations with the $p$-states of the N atoms (thus demonstrating the benefitted of the $d$-$p$ hopping integral). Therefore, they have dominant contribution in specifying the $J_1$ matrix (see Fig. AV(a-c)). In the CuN$_4$-G (R1, R2, R3) layers, the electronic states above the Fermi level are mainly contributed by the $d_{xy}$-orbital of the Cu atom (see Fig. AV(d-f)) and this orbital has a significant contribution to the $J_1$ matrix. Although the integrated area of the CuN$_4$-G (R2) is more localized, it has dominant contribution in exchange interaction in comparison with the MnN$_4$-G (R2), FeN$_4$-G (R2), and CoN$_4$-G (R2) layers (see Fig. AV(g)). To explain the observed trend, we have calculated the energy difference of the $d$-orbitals of the M atoms and the $p$-orbitals of the N atoms. The obtained results show that the CuN$_4$-G (R2) layer has a lower energy difference than the others. This leads to the larger value of the $d$-$p$ hopping integral (see Table III). Differing from the other layers in the second category, only in CuN$_4$-G (R1, R2, R3) layers the localized $d_{xy}$-orbital (the only available unoccupied orbital) contributes to the $d$-$p$ hopping integral. Moreover, although the integrated areas in CuN$_4$-G (R2) layer is larger than those in the CuN$_4$-G (R1, R3) layers, as shown in Fig. AV(d-f), the exchange interaction $J_1$ is larger for CuN$_4$-G (R1) in comparison with the CuN$_4$-G (R2, R3) layers. However, the energy difference between the $d$-orbitals of the Cu atoms and the $p$-orbitals of the N atoms shows a similar trend for these layers (Table III). Therefore, it can be concluded that the shorter spacer in the CuN$_4$-G (R1) layer is clear evidence for strong superexchange interaction.

**Table III.** Energy difference (minority-spin channel) between the $d$- and $p$-orbitals for MnN$_4$-G (R2), FeN$_4$-G (R2), CoN$_4$-G (R2), and CuN$_4$-G (R1, R2, R3) layers.

| Layer | $\Delta E^{\downarrow}_{d_{yz}, p_z}$ (eV) | $\Delta E^{\downarrow}_{d_{yz}, p_x}$ (eV) | $\Delta E^{\downarrow}_{d_{yz}, p_y}$ (eV) | Layer | $\Delta E^{\downarrow}_{d_{xy}, p_z}$ (eV) | $\Delta E^{\downarrow}_{d_{xy}, p_x}$ (eV) | $\Delta E^{\downarrow}_{d_{xy}, p_y}$ (eV) |
|---|---|---|---|---|---|---|---|
| MnN$_4$-G (R2) | 6.32 | 5.75 | 5.23 | CuN$_4$-G (R1) | -0.65 | -1.29 | -1.69 |
| FeN$_4$-G (R2) | 5.53 | 4.87 | 4.50 | CuN$_4$-G (R2) | -0.61 | -1.15 | -1.61 |
| CoN$_4$-G (R2) | 5.44 | 4.96 | 4.42 | CuN$_4$-G (R3) | -0.57 | -1.05 | -1.46 |

Aside from the exchange interaction type, to confirm the long-range behavior of the exchange interactions in the first group, we fitted our data points of the exchange interaction along $x$, $y$, **d1** and **d2** directions with the following oscillatory function:

$$J_r = J_0 \left( \frac{Cos(2ak_F r)}{(2ak_F r)^n} \right), \qquad (5)$$

where $k_F$ is the Fermi wave vector, $J_0$ is a constant with a unit of energy, $a$ is the lattice constant along the $x$ direction, and $n$ shows the decay rate of the RKKY interaction. In Fig. 8 and BII we have reported the fitted plots and parameters, respectively.

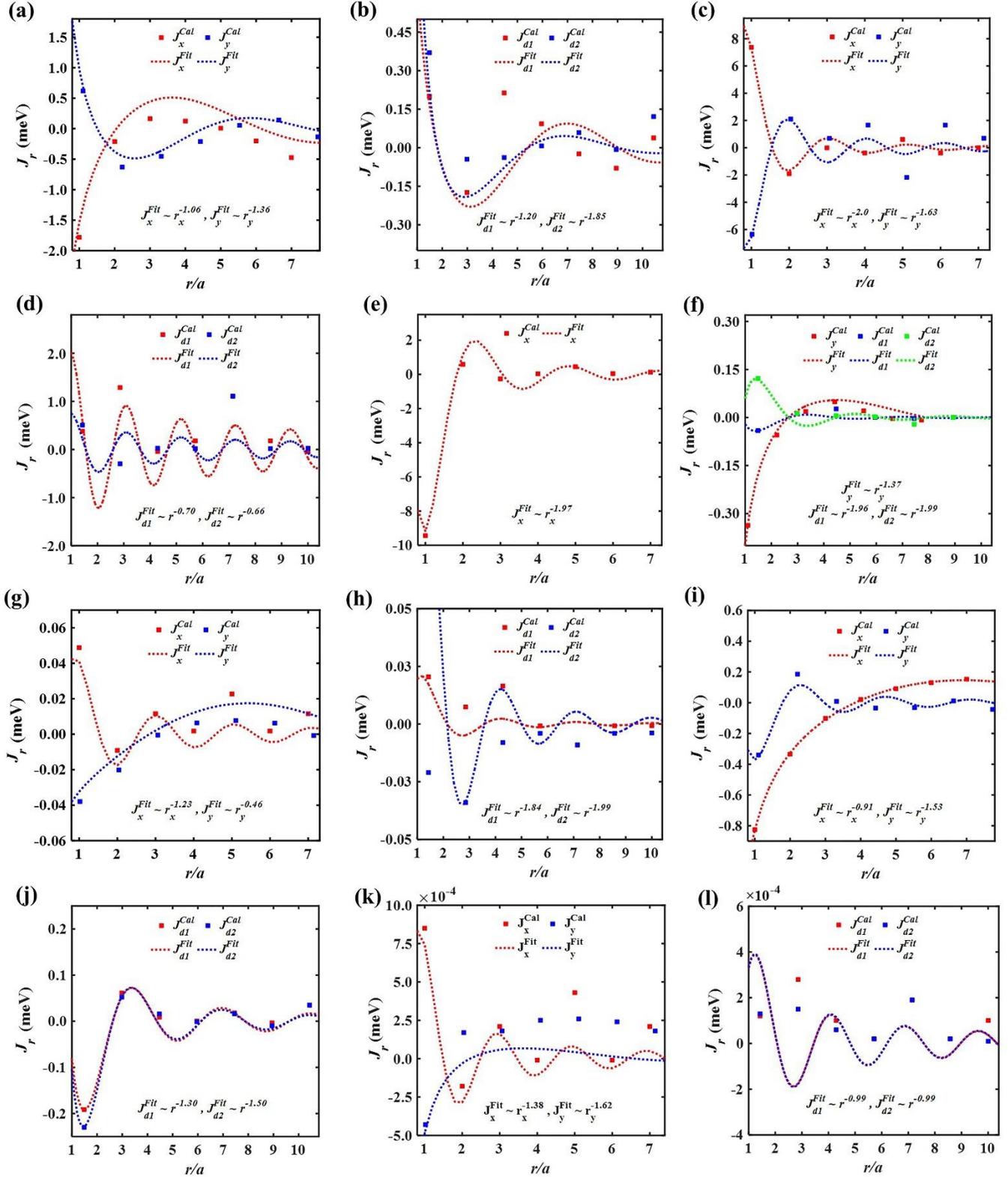

**FIG. 8.** Calculated ($J_r^{Cal}$) and fitted ($J_r^{Fit}$) exchange interactions as a function of M-M separation distance ($r$) along the $x$, $y$, and two main diameters (**d1**, **d2**) directions. $a$ is the lattice constant along the $x$ direction: (a, b) MnN$_4$-G (R1), (c, d) MnN$_4$-G (R3), (e, f) FeN$_4$-G (R1), (g, h) FeN$_4$-G (R3), (i, j) CoN$_4$-G (R1), and (k, l) CoN$_4$-G (R3) layers. The oscillations are dependent on the distance ($r$) and angle (with $x$-axis, $\theta$) that here only dependency on $r$ have shown.

The magnitude of the Fermi wave vector obtained by fitting is almost in agreement with our data (see Table BIII). In addition, the density plots of the calculated and fitted exchange interactions for MN$_4$-G (R1, R3) versus $r/a$ in different directions (Fig. 9) show both the fitting goodness and the compatibility of the calculated and fitted data.

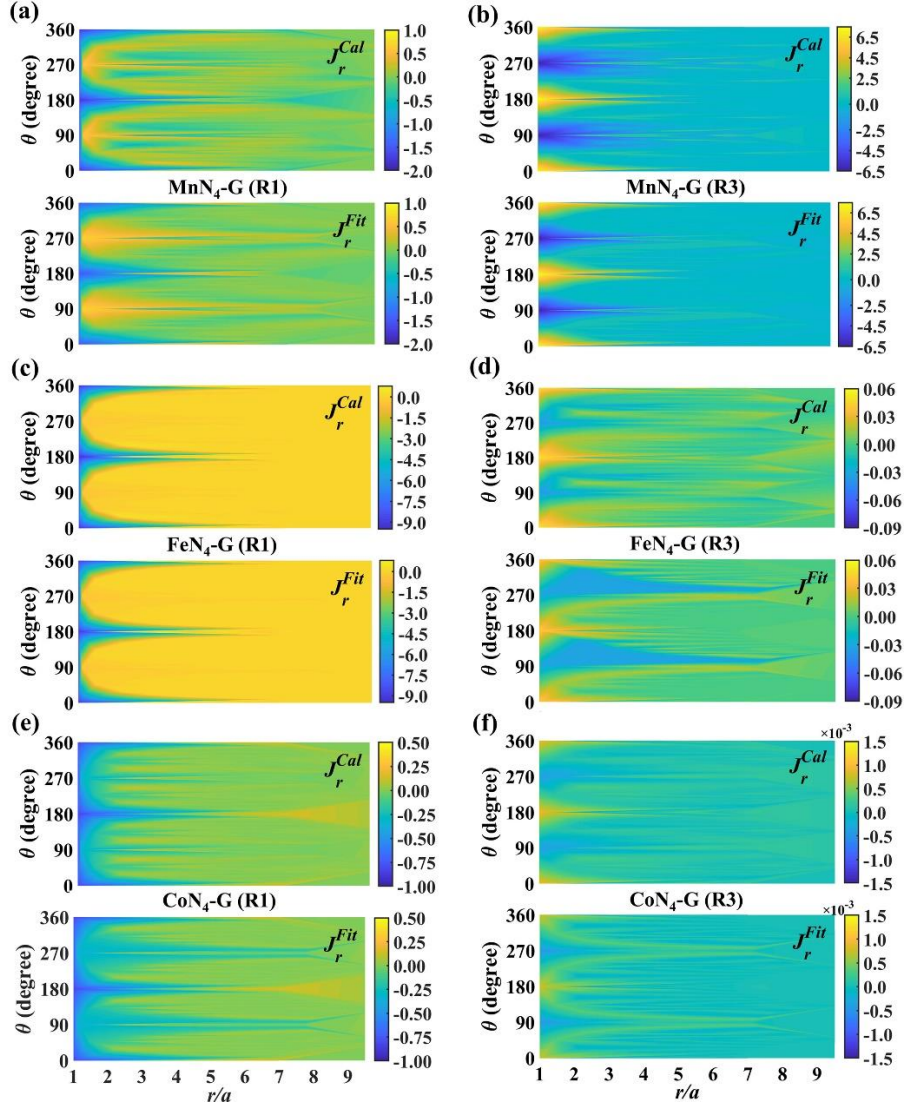

**FIG. 9.** The density plots of the calculated ($J_r^{Cal}$) and the fitted ($J_r^{Fit}$) exchange interactions for (a, b) MnN$_4$-G (R1, R3), (c, d) FeN$_4$-G (R1, R3), and (e, f) CoN$_4$-G (R1, R3) layers. Here $\theta$ is the angle of $r$ with $x$-axis.

## IV. CONCLUSIONS

In summary, having deep insight on the exchange mechanisms is essential for tuning the exchange interactions to suit the need of practical magnetic storage and spintronic devices. One of the main problems in this area is to understand how the chemical environments influence the exchange interactions. To this end, we have adopted the experimentally accessible MN$_4$-G layers with different sizes. Based on the dependence of the exchange interactions on the M-M separation distance, we have categorized the studied surfaces into two groups. The first group includes MnN$_4$-G (R1, R3), FeN$_4$-G (R1, R3), and CoN$_4$-G (R1, R3) layers that the exchange interaction is of RKKY type, with unusual slow decay rate in the form of $\frac{1}{r^{0.5 \leq n \leq 2}}$ which is an indication of the long-range interaction. The second group include the MnN$_4$-G (R2), FeN$_4$-G (R2), CoN$_4$-G (R2), and CuN$_4$-G (R1, R2, R3) layers, in which superexchange is the main mechanism responsible for the short-range exchange interaction. CuN$_4$-G is an exceptional case because all the related R1, R2, and R3 layers have exchange interaction in the $x$ direction and we can consider them as quasi-one-dimensional magnetic layers. The one-dimensional spin-1/2 chains, like CuN$_4$-G (R1, R2, R3), can be a representative of the Tomonaga–Luttinger liquid.

Our results provide deep insight into the magnetic properties of MN$_4$-G layers, and pave the road for studying other interesting magnetic properties like the effect of two types of magnetic atoms in supercells and spin dynamics that are on the way.

**Acknowledgements**

F. N. and J. A. are grateful to the Institute for Advanced Studies in Basic Sciences, for financial support through research grant No. **G2021IASBS32604** and **G2021IASBS12654**, respectively.

## APPENDIX A: DETAILES OF ELECTRONIC STRUCTURES

In this appendix we report the obtained results for spin-polarized projected density of states (PDOS) (Fig. AI), energy level splitting of the $d$-orbitals (Fig. AII), spin density distribution (Fig. AIII), charge density distribution (Fig. IV), integrated density of states for $d$-states (Fig. V),

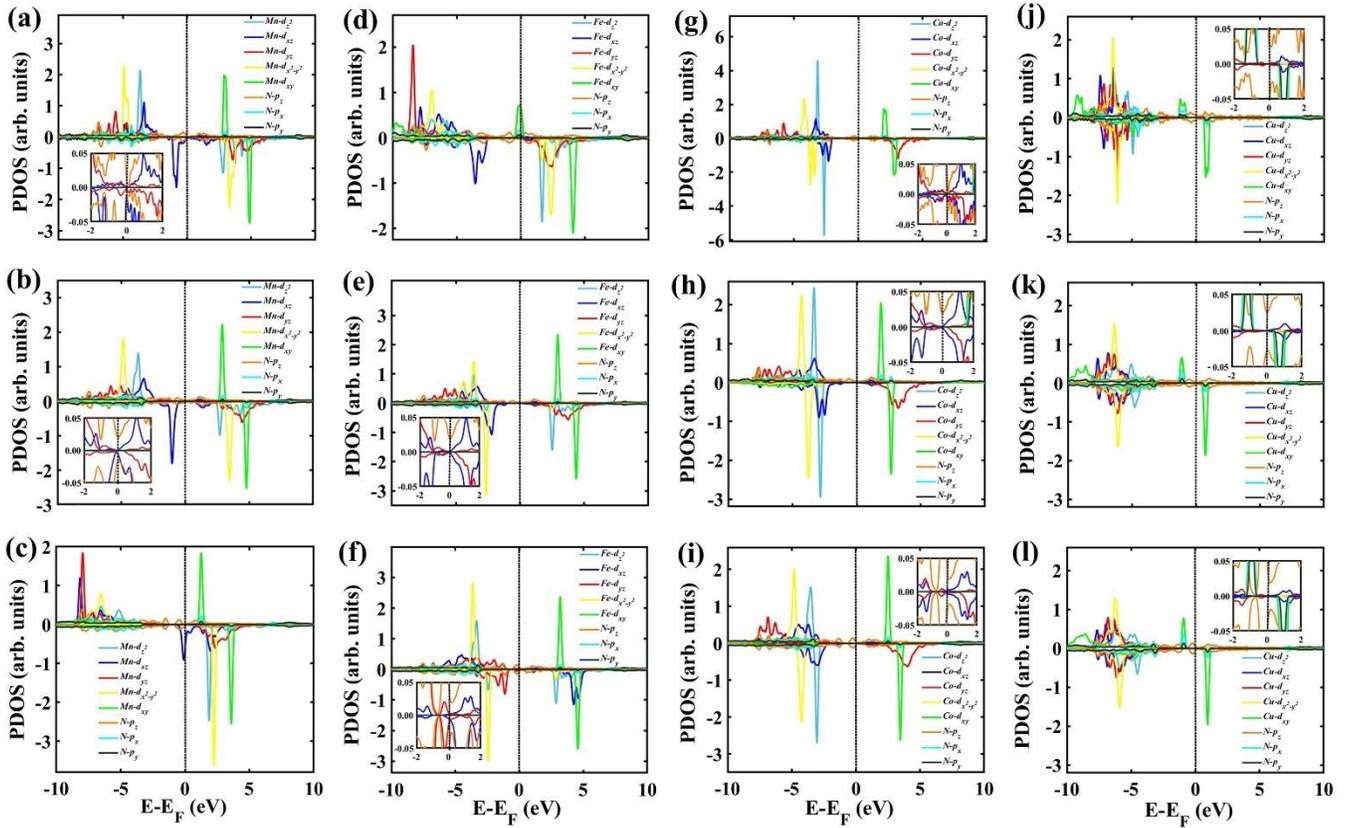

**FIG. AI.** Spin-polarized PDOS for (a) MnN$_4$-G (R1), (b) MnN$_4$-G (R2), (c) MnN$_4$-G (R3), (d) FeN$_4$-G (R1), (e) FeN$_4$-G (R2), (f) FeN$_4$-G (R3), (g) CoN$_4$-G (R1), (h) CoN$_4$-G (R2), (i) CoN$_4$-G (R3), (j) CuN$_4$-G (R1), (k) CuN$_4$-G (R2), and (l) CuN$_4$-G (R3) layers. The inset is a zoom of PDOS at low energy. The Fermi level is set to zero.

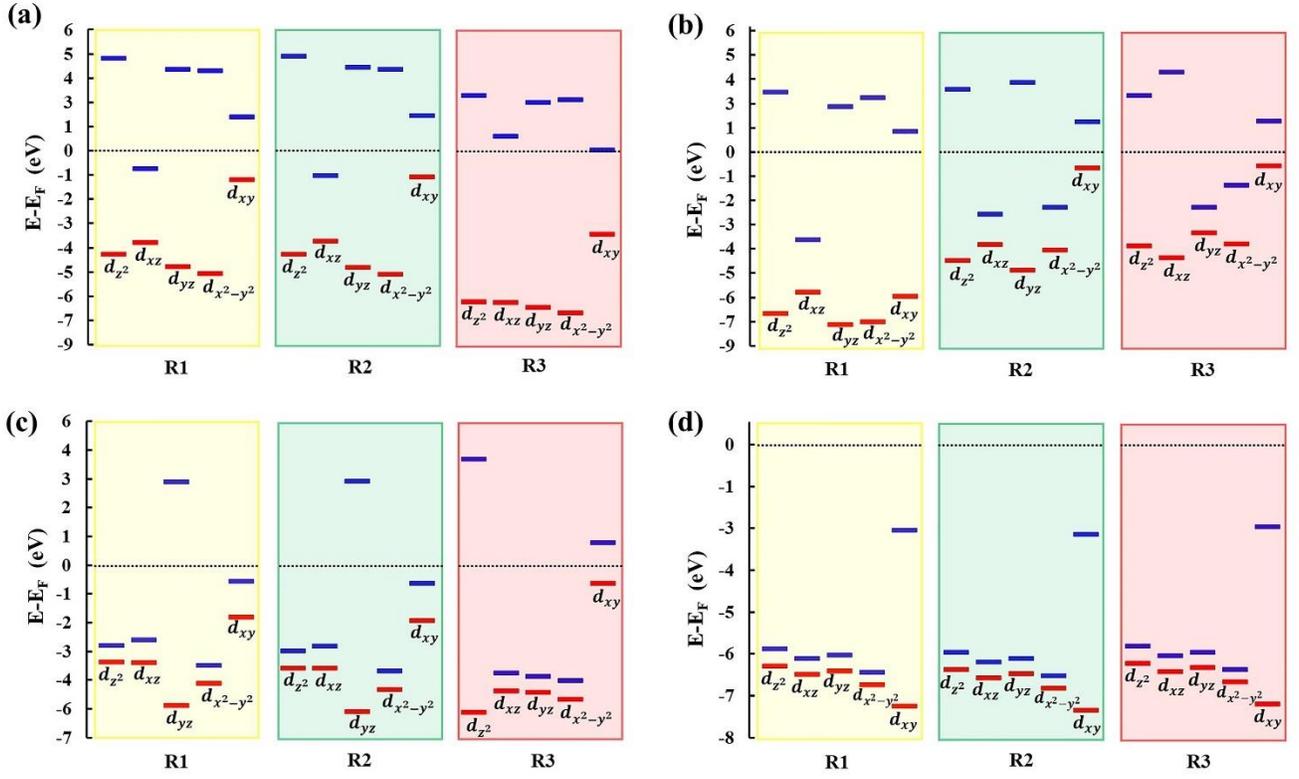

**FIG. AII.** Energy level splitting of the $d$-orbitals of M atoms for (a) MnN$_4$-G (R1, R2, R3), (b) FeN$_4$-G (R1, R2, R3), (c) CoN$_4$-G (R1, R2, R3), and (d) CuN$_4$-G (R1, R2, R3) layers. Red and blue colors represent the majority- and minority-spin energy levels, respectively.

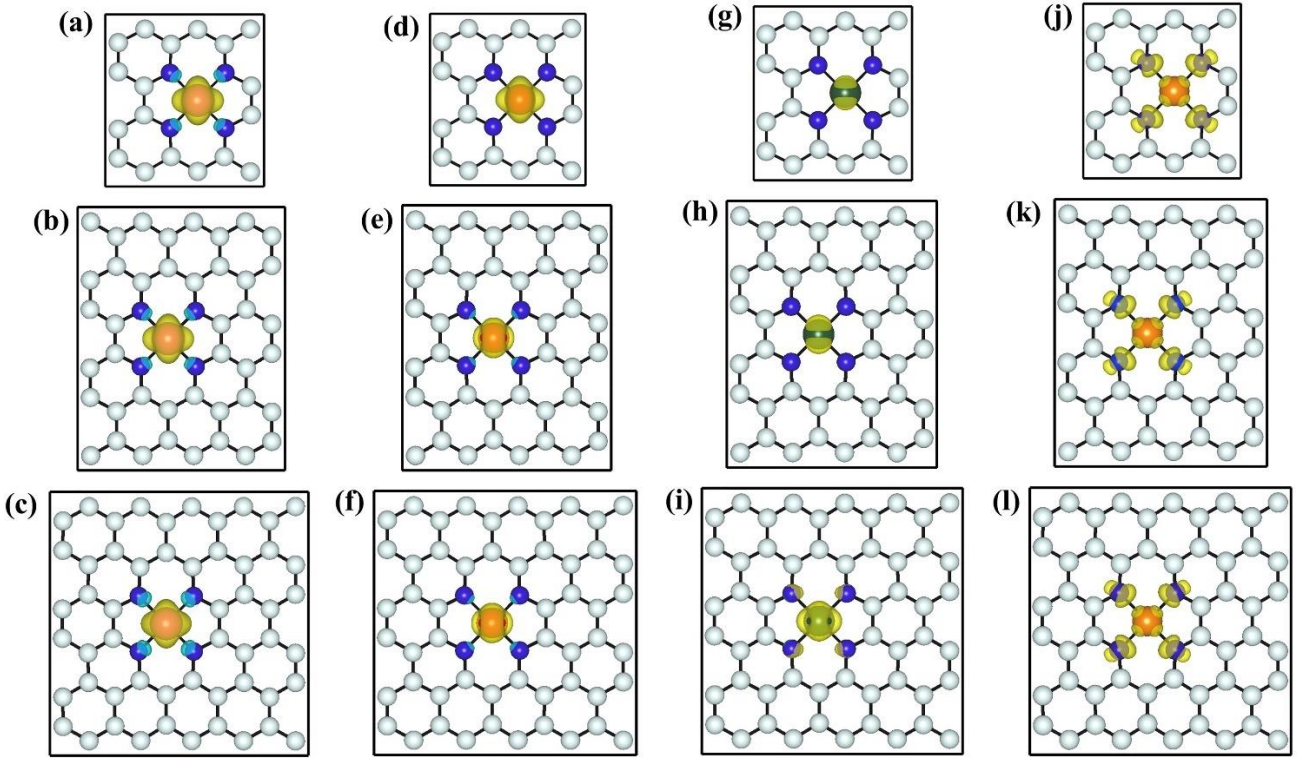

**FIG. AIII.** Spin density distribution (in the $xy$ plane) for (a) MnN$_4$-G (R1), (b) MnN$_4$-G (R2), (c) MnN$_4$-G (R3), (d) FeN$_4$-G (R1), (e) FeN$_4$-G (R2), (f) FeN$_4$-G (R3), (g) CoN$_4$-G (R1), (h) CoN$_4$-G (R2), (i) CoN$_4$-G (R3), (j) CuN$_4$-G (R1), (k) CuN$_4$-G (R2), and (l) CuN$_4$-G (R3) layers. The isosurface of spin density is set to 0.008 e/Å$^3$. Yellow and cyan colors indicate the majority- and minority-spin densities.

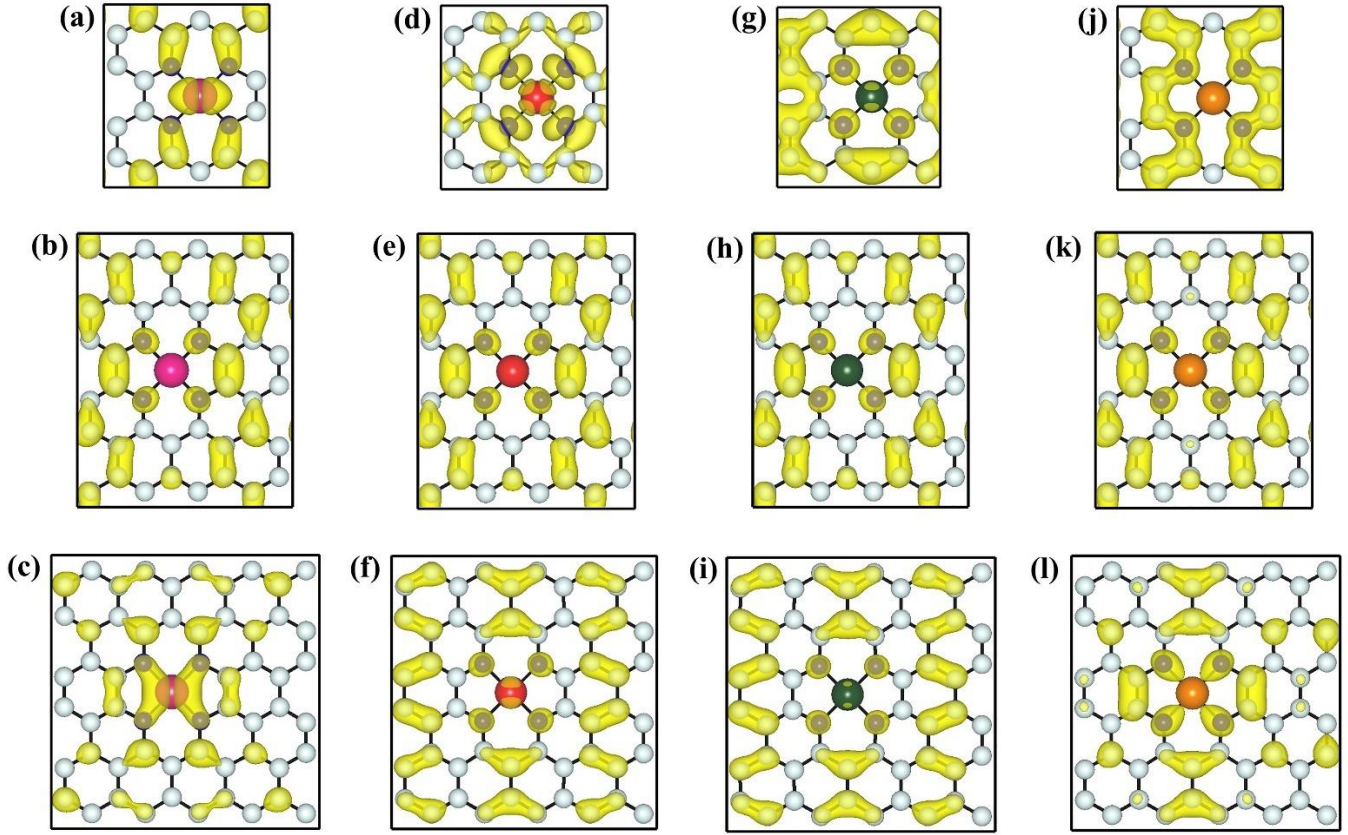

**FIG. AIV.** Charge density distribution (in the $xy$ plane) near the Fermi level for (a) MnN$_4$-G (R1), (b) MnN$_4$-G (R2), (c) MnN$_4$-G (R3), (d) FeN$_4$-G (R1), (e) FeN$_4$-G (R2), (f) FeN$_4$-G (R3), (g) CoN$_4$-G (R1), (h) CoN$_4$-G (R2), (i) CoN$_4$-G (R3), (j) CuN$_4$-G (R1), (k) CuN$_4$-G (R2), and (l) CuN$_4$-G (R3) layers. The isosurface of charge density is set to 0.0007 e/Å$^3$.

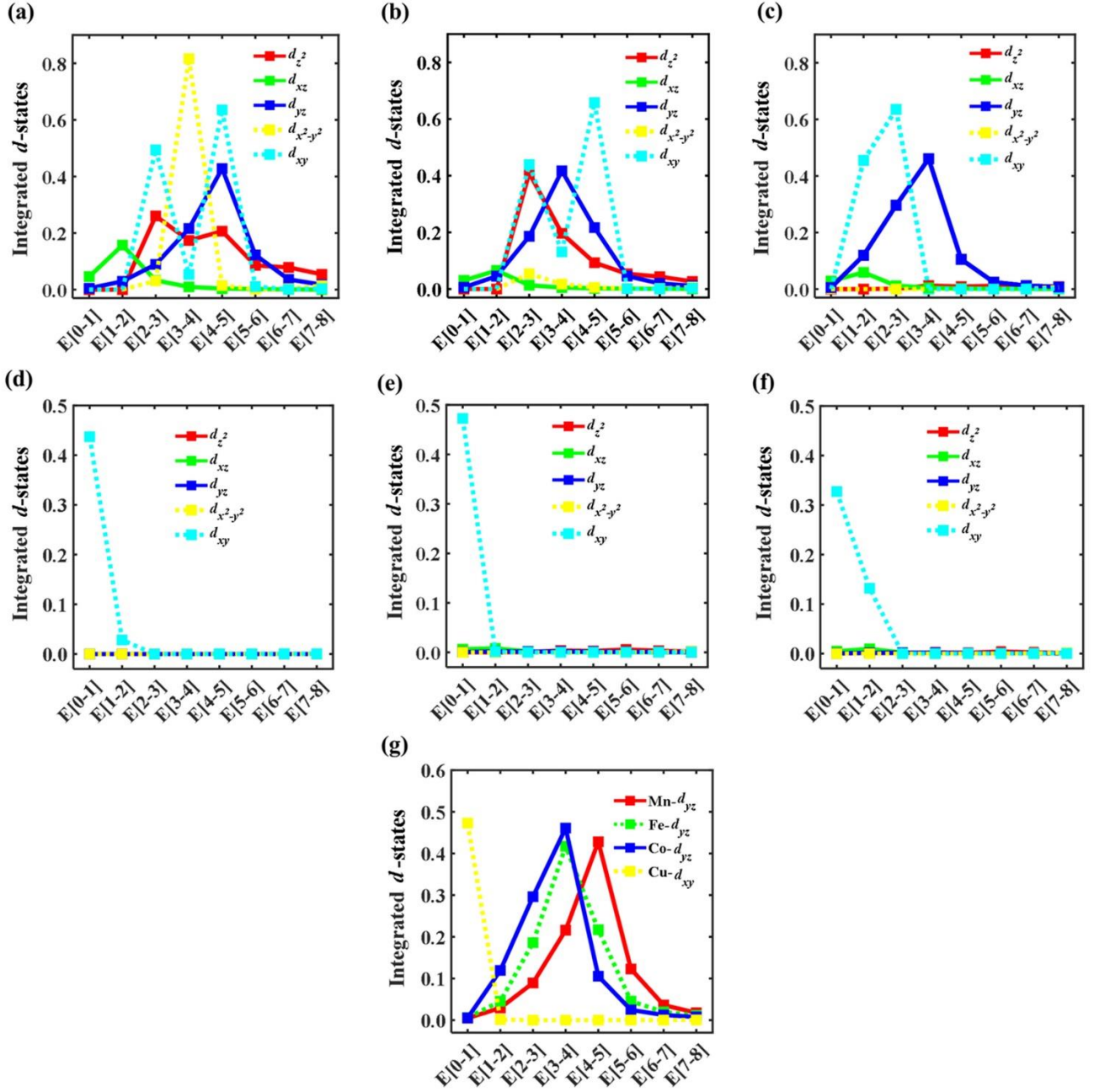

**FIG. AV.** Integrated $d$-states (sum of majority- and minority-spin $d$-states) verses E-E$_F$ in different energy levels for (a) MnN$_4$-G (R2), (b) FeN$_4$-G (R2), (c) CoN$_4$-G (R2), (d) CuN$_4$-G (R1), (e) CuN$_4$-G (R2), (f) CuN$_4$-G (R3). In (g) we plotted the integrated $d_{yz}$ states of MN$_4$-G (R2) with Mn, Fe, Co layers and $d_{xy}$ states of CuN$_4$-G (R2) in different energy intervals.

## APPENDIX B: DETAILES OF EXCHNAGE INTERACTION

In this appendix, we have reported comparison of the orbital contribution to exchange interaction $J_1$ (Fig. BI). In addition, fitting parameters for $J_r$ in Eq. 5, Fermi wave vector and its components are summarized in Tables BII and B III, respectively.

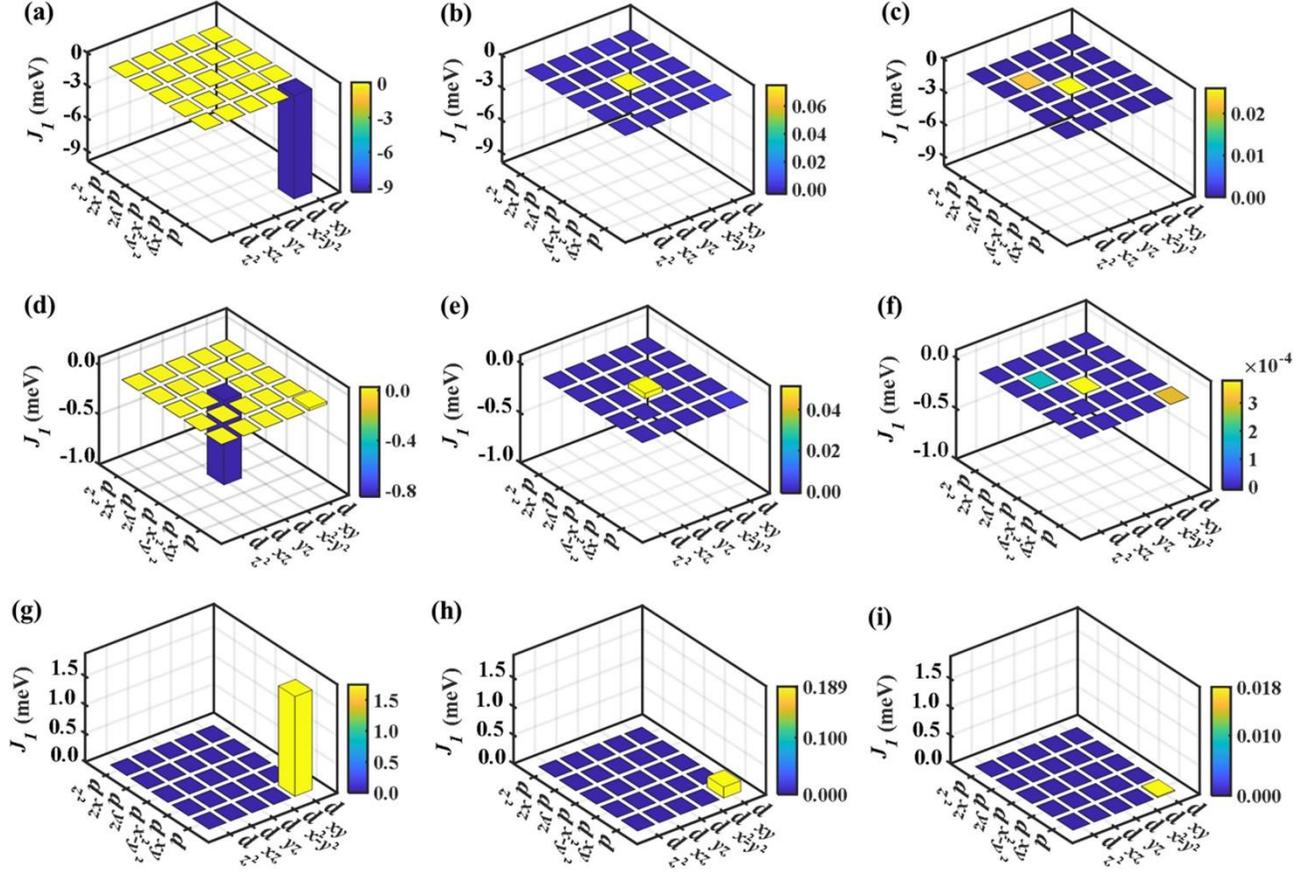

**FIG. BI.** $J_1$ matrix elements of (a) FeN$_4$-G (R1), (b) FeN$_4$-G (R2), (c) FeN$_4$-G (R3), (d) CoN$_4$-G (R1), (e) CoN$_4$-G (R2), (f) CoN$_4$-G (R3), (g) CuN$_4$-G (R1), (h) CuN$_4$-G (R2), and (i) CuN$_4$-G (R3) layers.

**Table BII.** Fitting parameters of $J_r$ in Eq. 5, for MN$_4$-G (R1, R3) layers, where $k_F$ is Fermi wave vector, $k_F^x$ and $k_F^y$ are the component of the Fermi wave vector; $J_0$ is a constant, and $n$ is decay rate of the RKKY interaction.

| Layer | Direction | $J_0$ (meV) | $k_F^x$ (Å$^{-1}$) | $k_F^y$ (Å$^{-1}$) | $k_F$ (Å$^{-1}$) | $n$ |
|---|---|---|---|---|---|---|
| MnN$_4$-G (R1) | x | -1.609 | 0.051 | ------ | ------ | 1.06 |
|  | y | 2.070 | ------ | 0.070 | ------ | 1.36 |
|  | d1 | 0.837 | ------ | ------ | 0.087 | 1.20 |
|  | d2 | 1.306 | ------ | ------ | 0.087 | 1.85 |
| MnN$_4$-G (R3) | x | -75.610 | 0.123 | ------ | ------ | 2.00 |
|  | y | 41.040 | ------ | 0.125 | ------ | 1.63 |
|  | d1 | -4.402 | ------ | ------ | 0.175 | 0.70 |
|  | d2 | -1.570 | ------ | ------ | 0.175 | 0.66 |
| FeN$_4$-G (R1) | x | 70.00 | 0.172 | ------ | ------ | 1.97 |
|  | y | -0.236 | ------ | 0.040 | ------ | 1.37 |
|  | d1 | 0.315 | ------ | ------ | 0.177 | 1.96 |
|  | d2 | -0.906 | ------ | ------ | 0.177 | 1.99 |
| FeN$_4$-G (R3) | x | -0.164 | 0.125 | ------ | ------ | 1.23 |
|  | y | -0.029 | ------ | 0.022 | ------ | 0.46 |
|  | d1 | -0.143 | ------ | ------ | 0.127 | 1.84 |
|  | d2 | -1.300 | ------ | ------ | 0.127 | 1.99 |
| CoN$_4$-G (R1) | x | -0.397 | 0.027 | ------ | ------ | 0.91 |
|  | y | 1.859 | ------ | 0.177 | ------ | 1.53 |
|  | d1 | 0.772 | ------ | ------ | 0.179 | 1.30 |
|  | d2 | 1.132 | ------ | ------ | 0.179 | 1.50 |
| CoN$_4$-G (R3) | x | -0.004 | 0.129 | ------ | ------ | 1.38 |
|  | y | -0.0004 | ------ | 0.029 | ------ | 1.62 |
|  | d1 | -0.001 | ------ | ------ | 0.132 | 0.99 |
|  | d2 | -0.001 | ------ | ------ | 0.132 | 0.99 |

**Table BIII.** Calculated and fitted Fermi wave vectors for MN$_4$-G (R1, R3) layers.

| Layer | $k_F^{Cal}$ (Å$^{-1}$) | $k_F^{Fit}$ (Å$^{-1}$) |
|---|---|---|
| MnN$_4$-G (R1) | 0.089 | 0.087 |
| MnN$_4$-G (R3) | 0.125 | 0.175 |
| FeN$_4$-G (R1) | 0.198 | 0.177 |
| FeN$_4$-G (R3) | 0.120 | 0.127 |
| CoN$_4$-G (R1) | 0.135 | 0.179 |
| CoN$_4$-G (R3) | 0.105 | 0.132 |

# References

[1] J. Girovsky, J. Nowakowski, M. E. Ali, M. Baljozovic, H. R. Rossmann, T. Nijs, E. A. Aeby, S. Nowakowska, D. Siewert, G. Srivastava *et al.*, Long-range ferrimagnetic order in a two-dimensional supramolecular Kondo lattice, Nat. Commun. **8**, 1 (2017).

[2] B. Shabbir, M. Nadeem, Z. Dai, M. S. Fuhrer, Q. K. Xue, X. Wang, and Q. Bao, Long range intrinsic ferromagnetism in two dimensional materials and dissipationless future technologies, Appl. Phys. Rev. **5**, 041105 (2018).

[3] P. Huang, P. Zhang, S. Xu, H. Wang, X. Zhang, and H. Zhang, Recent advances in two-dimensional ferromagnetism: materials synthesis, physical properties and device applications, Nanoscale **12**, 2309 (2020).

[4] Y. Wang and J. Yi, Ferromagnetism in two-dimensional materials via doping and defect engineering, *In Spintronic 2D Materials,* edited by W. Liu and Y. Xu, 1st ed. (Elsevier, 2020).

[5] C. Gong, L. Li, Z. Li, H. Ji, A. Stern, Y. Xia, T. Cao, W. Bao, C. Wang, Y. Wang *et al.*, Discovery of intrinsic ferromagnetism in two-dimensional van der Waals crystals, Nature **546**, 265 (2017).

[6] M. Steinbrecher, R. Rausch, K. T. That, J. Hermenau, A. A. Khajetoorians, M. Potthoff, R. Wiesendanger, and J. Wiebe, Non-collinear spin states in bottom-up fabricated atomic chains, Nat. Commun. **9**, 1 (2018).

[7] M. A. H. Vozmediano, M. P. López-Sancho, T. Stauber, and F. Guinea, Local defects and ferromagnetism in graphene layers, Phys. Rev. B **72**, 155121 (2005).

[8] B. Wunsch, T. Stauber, F. Sols, and F. Guinea, Dynamical polarization of graphene at finite doping, New J. Phys. **8**, 318 (2006).

[9] S. Saremi, RKKY in half-filled bipartite lattices: Graphene as an example, Phys. Rev. B **76**, 184430 (2007).

[10] D. F. Kirwan, C. G. Rocha, A. T. Costa, and M. S. Ferreira, Sudden decay of indirect exchange coupling between magnetic atoms on carbon nanotubes, Phys. Rev. B **77**, 085432 (2008).

[11] M. Sherafati and S. Satpathy, RKKY interaction in graphene from the lattice Green's function, Phys. Rev. B **83**, 165425 (2011).

[12] A. M. Black-Schaffer, RKKY coupling in graphene, Phys. Rev. B **81**, 205416 (2010).

[13] B. Uchoa, T. G. Rappoport, and A. H. C. Neto, Kondo quantum criticality of magnetic adatoms in graphene, Phys. Rev. Lett. **106**, 016801 (2011).

[14] R. Singh and P. Kroll, Magnetism in graphene due to single-atom defects: dependence on the concentration and packing geometry of defects, J. Phys.: Condens. Matter **21**, 196002 (2009).

[15] E. J. G. Santos, D. Sánchez-Portal, and A. Ayuela, Magnetism of substitutional Co impurities in graphene: Realization of single π vacancies, Phys. Rev. B **81**, 125433 (2010).

[16] A. M. Black-Schaffer, Importance of electron-electron interactions in the RKKY coupling in graphene, Phys. Rev. B **82**, 073409 (2010).

[17] H. Lee, J. Kim, E. R. Mucciolo, G. Bouzerar, and S. Kettemann, RKKY interaction in disordered graphene, Phys. Rev. B **85**, 075420 (2012).

[18] E. Kogan, RKKY interaction in graphene, Phys. Rev. B **84**, 115119 (2011).

[19] L. Brey, H. A. Fertig, and S. D. Sarma, Diluted graphene antiferromagnet, Phys. Rev. Lett. **99**, 116802 (2007).

[20] D. Solenov, C. Junkermeier, T. L. Reinecke, and K. A. Velizhanin, Tunable adsorbate-adsorbate interactions on graphene, Phys. Rev. Lett. **111**, 115502 (2013).

[21] H. Park, A. Wadehra, J. W. Wilkins, and A. H. C. Neto, Spin-polarized electronic current induced by sublattice engineering of graphene sheets with boron/nitrogen, Phys. Rev. B **87**, 085441 (2013).

[22] S. R. Power and M. S. Ferreira, Indirect exchange and Ruderman–Kittel–Kasuya–Yosida (RKKY) interactions in magnetically-doped graphene, Crystals **3**, 49 (2013).

[23] M. Sherafati and S. Satpathy, Analytical expression for the RKKY interaction in doped graphene, Phys. Rev. B **84**, 125416 (2011).

[24] J. M. Duffy, P. D. Gorman, S. R. Power, and M. S. Ferreira, Variable range of the RKKY interaction in edged graphene, J. Phys.: Condens. Matter. **26**, 055007 (2013).

[25] K. Szałowski, Indirect RKKY interaction between localized magnetic moments in armchair graphene nanoribbons, J. Phys.: Condens. Matter. **25**, 166001 (2013).

[26] J. Klinovaja and D. Loss, RKKY interaction in carbon nanotubes and graphene nanoribbons, Phys. Rev. B **87**, 045422 (2013).

[27] V. V. Cheianov and V. I. Fal'ko, Friedel oscillations, impurity scattering, and temperature dependence of resistivity in graphene, Phys. Rev. Lett. **97**, 226801 (2006).

[28] A. V. Shytov, D. A. Abanin, and L. S. Levitov, Long-range interaction between adatoms in graphene, Phys. Rev. Lett. **103**, 016806 (2009).

[29] J. Ren, H. Guo, J. Pan, Y. F. Zhang, Y. Yang, X. Wu, S. Du, M. Ouyang, and H. J. Gao, Interatomic spin coupling in manganese clusters registered on graphene, Phys. Rev. Lett. **119**, 176806 (2017).

[30] Y. Zhu, Y. F. Pan, Z. Q. Yang, X. Y. Wei, J. Hu, Y. P. Feng, H. Zhang, and R. Q. Wu, Ruderman–Kittel–Kasuya–Yosida Mechanism for Magnetic Ordering of

Sparse Fe Adatoms on Graphene, J. Phys. Chem. C **123**, 4441 (2019).

[31] P. D. Gorman, J. M. Duffy, S. R. Power, and M. S. Ferreira, RKKY interaction between extended magnetic defect lines in graphene, Phys. Rev. B **90**, 125411 (2014).

[32] Y. P. Feng, L. Shen, M. Yang, A. Wang, M. Zeng, Q. Wu, S. Chintalapati, and C. R. Chang, Prospects of spintronics based on 2D materials, Wiley Interdiscip. Rev. Comput. Mol. Sci. **7**, e1313 (2017).

[33] I. Choudhuri, P. Bhauriyal, and B. Pathak, Recent advances in graphene-like 2D materials for spintronics applications, Chem. Mater. **31**, 8260 (2019).

[34] S. Wei1, X. Tang, X. Liao, Y. Ge, H. Jin, W. Chen, H. Zhang, and Y. Wei, Recent progress of spintronics based on emerging 2D materials: $CrI_3$ and Xenes, Mater. Res. Express **6**, 122004 (2019).

[35] Y. Liu, C. Zeng, J. Zhong, J. Ding, Z. M. Wang, and Z. Liu, Spintronics in two-dimensional materials, Nano-Micro Lett. **12**, 1 (2020).

[36] R. Langer, K. Mustonen, A. Markevich, M. Otyepka, T. Susi, and P. Błoński, Graphene Lattices with Embedded Transition-Metal Atoms and Tunable Magnetic Anisotropy Energy: Implications for Spintronic Devices, ACS Appl. Nano Mater. **5**, 1562 (2022).

[37] N. Tombros, C. Jozsa, M. Popinciuc, H. T. Jonkman, and B. J. V. Wees, Electronic spin transport and spin precession in single graphene layers at room temperature, Nature **448**, 571 (2007).

[38] V. Georgakilas, J. A. Perman, J. Tucek, and R. Zboril, Broad family of carbon nanoallotropes: classification, chemistry, and applications of fullerenes, carbon dots, nanotubes, graphene, nanodiamonds, and combined superstructures, Chem. Rev. **115**, 4744 (2015).

[39] O. V. Yazyev, Magnetism in disordered graphene and irradiated graphite, Phys. Rev. Lett. **101**, 037203 (2008).

[40] P. Esquinazi, D. Spemann, R. Höhne, A. Setzer, K. H. Han, and T. Butz, Induced magnetic ordering by proton irradiation in graphite, Phys. Rev. Lett. **91**, 227201 (2003).

[41] F. Ostovari, M. Hasanpoori, M. Abbasnejad, M. A. Salehi, DFT calculations of graphene monolayer in presence of Fe dopant and vacancy, Phys. B: Condens. Matter **541**, 6 (2018).

[42] R. Sielemann, Y. Kobayashi, Y. Yoshida, H. P. Gunnlaugsson, and G. Weyer, Magnetism at single isolated iron atoms implanted in graphite, Phys. Rev. Lett. **101**, 137206 (2008).

[43] A. V. Krasheninnikov, P. O. Lehtinen, A. S. Foster, P. Pyykkö, and R. M. Nieminen, Embedding transition-metal atoms in graphene: structure, bonding, and magnetism, Phys. Rev. Lett. **102**, 126807 (2009).

[44] J. Tuček, P. Błoński, Z. Sofer, P. Šimek, M. Petr, M. Pumera, M. Otyepka, and R. Zbořil, Sulfur doping induces strong ferromagnetic ordering in graphene: effect of concentration and substitution mechanism, Adv. Mater. **28**, 5045 (2016).

[45] J. J. Palacios, J. Fernández-Rossier, and L. Brey, Vacancy-induced magnetism in graphene and graphene ribbons, Phys. Rev. B **77**, 195428 (2008).

[46] J. Tuček, K. Holá, A. B. Bourlinos, P. Błoński, A. Bakandritsos, J. Ugolotti, M. Dubecký, F. Karlický, V. Ranc, K. Čépe et al., Room temperature organic magnets derived from $sp^3$ functionalized graphene, Nat. Commun. **8**, 1 (2017).

[47] P. Błoński, J. Tuček, Z. Sofer, V. Mazánek, M. Petr, M. Pumera, M. Otyepka, and R. Zbořil, Doping with graphitic nitrogen triggers ferromagnetism in graphene, J. Am. Chem. Soc. **139**, 3171 (2017).

[48] H. G. Herrero, J. M. G. Rodríguez, P. Mallet, M. Moaied, J. J. Palacios, C. Salgado, M. M. Ugeda, J. Y. Veuillen, F. Yndurain, and I. Brihuega, Atomic-scale control of graphene magnetism by using hydrogen atoms, Science **352**, 437 (2016).

[49] G. Z. Magda, X. Jin, I. Hagymási, P. Vancsó, Z. Osváth, P. N. Incze, C. Hwang, L. P. Biró, and L. Tapasztó, Room-temperature magnetic order on zigzag edges of narrow graphene nanoribbons, Nature **514**, 608 (2014).

[50] Y. Ito, C. Christodoulou, M. V. Nardi, N. Koch, M. Kläui, H. Sachdev, and K. Müllen, Tuning the magnetic properties of carbon by nitrogen doping of its graphene domains, J. Am. Chem. Soc. **137**, 7678 (2015).

[51] R. R. Nair, M. Sepioni, I. L. Tsai, O. Lehtinen, J. Keinonen, A. V. Krasheninnikov, T. Thomson, A. K. Geim, and I. V. Grigorieva, Spin-half paramagnetism in graphene induced by point defects, Nat. Phys. **8**, 199 (2012).

[52] Y. Liu, Y. Shen, L. Sun, J. Li, C. Liu, W. Ren, F. Li, L. Gao, J. Chen, F. Liu et al., Elemental superdoping of graphene and carbon nanotubes, Nat. Commun. **7**, 1 (2016).

[53] W. Hu, C. Wang, H. Tan, H. Duan, G. Li, N. Li, Q. Ji, Y. Lu, Y. Wang, Z. Sun et al., Embedding atomic cobalt into graphene lattices to activate room-temperature ferromagnetism, Nat. Commun. **12**, 1 (2021).

[54] N. D. Mermin and H. Wagner, Absence of ferromagnetism or antiferromagnetism in one-or two-dimensional isotropic Heisenberg models, Phys. Rev. Lett. **17**, 1133 (1966).

[55] A. C. Hewson, *The Kondo problem to heavy fermions* (Cambridge university press, 1993).

[56] B. Xia, Z. Liao, Y. Liu, X. Chi, W. Xiao, J. Ding, T. Wang, D. Gao, and D. Xue, Realization of "single-atom ferromagnetism" in graphene by Cu–$N_4$ moieties anchoring, Appl. Phys. Lett. **116**, 113102 (2020).

[57] H. Zhang, Z. Liao, B. Xia, T. S. Herng, J. Ding, and D. Gao, Ferromagnetism of Mn-$N_4$ Architecture Embedded Graphene, J. Phys. D: Appl. Phys. **55**, 225001 (2022).

[58] M. Garnica, D. Stradi, S. Barja, F. Calleja, C. Díaz, M. Alcamí, N. Martín, A. L. Vázquez de Parga, F. Martín et al., Long-range magnetic order in a purely organic

2D layer adsorbed on epitaxial graphene, Nat. Phys. **9**, 368 (2013).

[59] A. M. Tokmachev, D. V. Averyanov, O. E. Parfenov, A. N. Taldenkov, I. A. Karateev, I. S. Sokolov, O. A. Kondratev, and V. G. Storchak, Emerging two-dimensional ferromagnetism in silicene materials, Nat. Commun. **9**, 1 (2018).

[60] R. Tuerhong, F. Ngassam, S. Watanabe, J. Onoe, M. Alouani, and J. P. Bucher, Two-dimensional organometallic Kondo lattice with long-range antiferromagnetic order, J. Phys. Chem. C **122**, 20046 (2018).

[61] A. I. Liechtenstein, V. A. Gubanov, M. I. Katsnelson, and V. I. Anisimov, Magnetic transition state approach to ferromagnetism of metals: Ni, J. Magn. Magn. Mater. **36**, 125 (1983).

[62] A. I. Liechtenstein, M. I. Katsnelson, V. P. Antropov, and V. A. Gubanov, Local spin density functional approach to the theory of exchange interactions in ferromagnetic metals and alloys, J. Magn. Magn. Mater. **67**, 65 (1987).

[63] M. I. Katsnelson and A. I. Lichtenstein, First-principles calculations of magnetic interactions in correlated systems, Phys. Rev. B **61**, 8906 (2000).

[64] V. I. Anisimov, D. E. Kondakov, A. V. Kozhevnikov, I. A. Nekrasov, Z. V. Pchelkina, J. W. Allen, S. K. Mo, H. D. Kim, P. Metcalf, S. Suga *et al.*, Full orbital calculation scheme for materials with strongly correlated electrons, Phys. Rev. B **71**, 125119 (2005).

[65] Dm. M. Korotin, A. V. Kozhevnikov, S. L. Skornyakov, I. Leonov, N. Binggeli, V. I. Anisimov, and G. Trimarchi, Construction and solution of a Wannier-functions based Hamiltonian in the pseudopotential plane-wave framework for strongly correlated materials, Eur. Phys. J. B **65**, 91 (2008).

[66] W. Orellana, Catalytic properties of transition metal–$N_4$ moieties in graphene for the oxygen reduction reaction: evidence of spin-dependent mechanisms, J. Phys. Chem. C **117**, 9812 (2013).

[67] K. Liu, G. Wu, and G. Wang, Role of local carbon structure surrounding $FeN_4$ sites in boosting the catalytic activity for oxygen reduction, J. Phys. Chem. C **121**, 11319 (2017).

[68] D. Deng, X. Chen, L. Yu, X. Wu, Q. Liu, Y. Liu, H. Yang, H. Tian, Y. Hu, P. Du *et al.*, A single iron site confined in a graphene matrix for the catalytic oxidation of benzene at room temperature, Sci. adv. **1**, e1500462 (2015).

[69] T. Patniboon and H. A. Hansen, Acid-Stable and Active M–N–C Catalysts for the Oxygen Reduction Reaction: The Role of Local Structure, ACS Catal. **11**, 13102 (2021).

[70] L. Wu, X. Cao, W. Hu, Y. Ji, Z. Z. Zhu, and X. F. Li, Improving the oxygen reduction reaction activity of $FeN_4$–graphene via tuning electronic characteristics, ACS Appl. Energy Mater. **2**, 6634 (2019).

[71] E. Ashori, F. Nazari, and F. Illas, Influence of NO and $(NO)_2$ adsorption on the properties of Fe-$N_4$ porphyrin-like graphene sheets, Phys. Chem. Chem. Phys. **19**, 3201 (2017).

[72] T. Giamarchi, *Quantum physics in one dimension* (Clarendon press, 2003).

[73] P. Giannozzi, S. Baroni, N. Bonini, M. Calandra, R. Car, C. Cavazzoni, D. Ceresoli, G. L. Chiarotti, M. Cococcioni, I. Dabo *et al.*, QUANTUM ESPRESSO: a modular and open-source software project for quantum simulations of materials, J. Phys.: Condens. Matter **21**, 395502 (2009).

[74] C. Fiolhais, F. Nogueira, and M. A. Marques, *A primer in density functional theory* (Springer Science & Business Media, 2003).

[75] J. P. Perdew, A. Ruzsinszky, G. I. Csonka, O. A. Vydrov, G. E. Scuseria, L. A. Constantin, X. Zhou, and K. Burke, Restoring the density-gradient expansion for exchange in solids and surfaces, Phys. Rev. Lett. **100**, 136406 (2008).

[76] J. P. Perdew, K. Burke, and M. Ernzerhof, Generalized gradient approximation made simple, Phys. Rev. Lett. **77**, 3865 (1996).

[77] P. E. Blöchl, Projector augmented-wave method, Phys. Rev. B **50**, 17953 (1994).

[78] H. J. Monkhorst and J. D. Pack, Special points for Brillouin-zone integrations, Phys. Rev. B **13**, 5188 (1976).

[79] I. Timrov, N. Marzari, and M. Cococcioni, Hubbard parameters from density-functional perturbation theory, Phys. Rev. B **98**, 085127 (2018).

[80] L. Noodleman, Valence bond description of antiferromagnetic coupling in transition metal dimers, J. Chem. Phys. **74**, 5737 (1981).

[81] A. A. Tsirlin, Spin-chain magnetism and uniform Dzyaloshinsky-Moriya anisotropy in $BaV_3O_8$, Phys. Rev. B **89**, 014405 (2014).

[82] D. Wu, Z. Zhuo, H. Lv, and X. Wu, Two-Dimensional $Cr_2X_3S_3$ (X= Br, I) Janus Semiconductor with Intrinsic Room-Temperature Magnetism, J. Phys. Chem. Lett. **12**, 2905 (2021).

[83] Dm. M. Korotin, V. V. Mazurenko, V. I. Anisimov, and S. V. Streltsov, Calculation of exchange constants of the Heisenberg model in plane-wave-based methods using the Green's function approach, Phys. Rev. B **91**, 224405 (2015).

[84] O. K. Andersen and O. Jepsen, Explicit, first-principles tight-binding theory, Phys. Rev. Lett. **53**, 2571 (1984).

[85] W. A. Harrison, *Elementary Electronic Structure* (World Scientific, Singapore, 1999).

[86] D. Singh, *Plane Waves, Pseudopotentials and the LAPW Method* (Kluwer Academic, The Netherlands, 1994).

[87] R. M. Martin, *Electronic Structure: Basic Theory and Practical Methods* (Cambridge University Press, Cambridge, 2004).

[88] M. Methfessel and J. Kubler, Bond analysis of heats of formation: application to some group VIII and IB hydrides, J. Phys. F: Met. Phys. **12**, 141 (1982).

[89] H. Yoon, T. J. Kim, J. H. Sim, S. W. Jang, T. Ozaki, and M. J. Han, Reliability and applicability of

magnetic-force linear response theory: Numerical parameters, predictability, and orbital resolution, Phys. Rev. B **97**, 125132 (2018).

[90] A. T. Lee, J. Kang, S. H. Wei, K. J. Chang, and Y. H. Kim, Carrier-mediated long-range ferromagnetism in electron-doped Fe-$C_4$ and Fe-$N_4$ incorporated graphene, Phys. Rev. B **86**, 165403 (2012).

[91] B. R. K. Nanda, M. Sherafati, Z. S. Popović, and S. Satpathy, Electronic structure of the substitutional vacancy in graphene: density-functional and Green's function studies, New J. Phys. **14**, 083004 (2012).

[92] A. Allerdt, H. Hafiz, B. Barbiellini, A. Bansil, and A. E. Feiguin, Many-body effects in $FeN_4$ center embedded in graphene, Appl. Sci. **10**, 2542 (2020).

[93] S. Aoyama, J. Kaiwa, P. Chantngarm, S. Tanibayashi, H. Saito, M. Hasegawa, and K. Nishidate, Oxygen reduction reaction of $FeN_4$ center embedded in graphene and carbon nanotube: Density functional calculations, AIP Adv. **8**, 115113 (2018).

[94] W. I. Choi, S. H. Jhi, K. Kim, and Y. H. Kim, Divacancy-nitrogen-assisted transition metal dispersion and hydrogen adsorption in defective graphene: A first-principles study, Phys. Rev. B **81**, 085441 (2010).

[95] M. Luo, Z. Liang, S. G. Peera, M. Chen, C. Liu, H. Yang, J. Liu, U. P. Kumar, and T. Liang, Theoretical study on the adsorption and predictive catalysis of $MnN_4$ embedded in carbon substrate for gas molecules, Appl. Surf. Sci. **525**, 146480 (2020).

[96] Q. Jia, N. Ramaswamy, H. Hafiz, U. Tylus, K. Strickland, G. Wu, B. Barbiellini, A. Bansil, E. F. Holby, P. Zelenay *et al.*, Experimental observation of redox-induced Fe–N switching behavior as a determinant role for oxygen reduction activity, ACS Nano **9**, 12496 (2015).

[97] G. J. Long and F. Grandjean, *Supermagnets, hard magnetic materials* (Springer Science & Business Media, 2012).

[98] P. S. Perlepe, D. Maniaki, E. Pilichos, E. Katsoulakou, and S. P. Perlepes, Smart ligands for efficient 3d-, 4d- and 5d-metal single-molecule magnets and single-ion magnets, Inorganics **8**, 39 (2020).

[99] S. Kattel, P. Atanassov, and B. Kiefer, Stability, electronic and magnetic properties of in-plane defects in graphene: a first-principles study, J. Phys. Chem. C **116**, 8161 (2012).

[100] T. Roongcharoen, S. Impeng, N. Kungwan, and S. Namuangruk, Revealing the effect of N-content in Fe doped graphene on its catalytic performance for direct oxidation of methane to methanol, Appl. Surf. Sci. **527**, 146833 (2020).

[101] M. Battocletti, H. Ebert, and H. Akai, Influence of gradient corrections to the local-density-approximation on the calculation of hyperfine fields in ferromagnetic Fe, Co, and Ni, Phys. Rev. B **53**, 9776 (1996).

[102] D. J. M. King, S. C. Middleburgh, P. A. Burr, T. M. Whiting, P. C. Fossati, and M. R. Wenman, Density functional theory study of the magnetic moment of solute Mn in bcc Fe, Phys. Rev. B **98**, 024418 (2018).

[103] S.R. Brankovic, K. Sendur, T. Klemmer, X. Yang, and E. C. Johns, *Magnetic Materials Processes and Devices VII and Electrodeposition of Alloys* (Electrochemical Society Proceeding Series, 2002).

[104] D. Khomskii, *Transition metal compounds* (Cambridge University Press, 2014).